\documentclass[12pt]{report}    

\usepackage{utdiss2}        

\usepackage{amsmath,amsthm,amsfonts,amscd}
\usepackage{subfigure}
\usepackage{amssymb}
\usepackage{psfrag}
\usepackage{graphicx}          

\author{Eric Alexander Nicholson}    

\address{15407 Wild Timber Trail\\ Cypress, Texas 77433}  

\title{Perturbative Wilsonian Formalism for Noncommutative Gauge Theories in the Matrix Representation}

\supervisor
    {Willy Fischler}

\committeemembers
    [Jacques Distler]
    [Vadim Kaplunovsky]
    [Sonia Paban]
    {Dan Freed}

\previousdegrees{B.S.}
\oneandonehalfspacequote

\theoremstyle{definition}

\theoremstyle{remark}

\newcommand{\latexe}{{\LaTeX\kern.125em2%
                      \lower.5ex\hbox{$\varepsilon$}}}

\chardef\bslash=`\\ 

\makeatletter       
\def\square{\RIfM@\bgroup\else$\bgroup\aftergroup$\fi
  \vcenter{\hrule\hbox{\vrule\@height.6em\kern.6em\vrule}%
                                              \hrule}\egroup}
\makeatother        

\makeindex    

\begin{document}

\copyrightpage          

%
%
%
\commcertpage           

\titlepage              

%
\begin{dedication}
\index{Dedication@\emph{Dedication}}%
Dedicated to my loving wife, Elizabeth, whose support has been
invaluable.
\end{dedication}

\begin{acknowledgments}     
\index{Acknowledgments@\emph{Acknowledgments}}%
I wish to thank all of the members of the Theory Group for
providing a stimulating scientific environment. However, Willy
Fischler and Sonia Paban are especially deserving of thanks for
all of the advice and support that they have provided over the
years. I would also like to thank Li Jiang for countless valuable
and spirited discussions.
\end{acknowledgments}

%
\utabstract
\index{Abstract}%
\indent We study the perturbative approach to the Wilsonian
integration of noncommutative gauge theories in the matrix
representation. We begin by motivating the study of noncommutative
gauge theories and reviewing the matrix formulation. We then
systematically develop the perturbative treatment of UV states and
calculate both the leading and next to leading order one- and
two-loop corrections to the quantum effective action. Throughout,
we discuss how our formalism clarifies problems associated with
UV-IR mixing, a particular emphasis being placed on the dipole
structure imposed by noncommutative gauge invariance. Ultimately,
using the structural understanding developed in this work, we are
able to determine the exact form of perturbative corrections in
the UV regime defined by $\theta\Lambda^2\gg 1$. Finally, we apply
our results to the analysis of the divergence structure and show
that $3+1$ and higher dimensional noncommutative theories that
allow renormalization beyond one-loop are not self-consistent.

\tableofcontents   

\listoffigures     

%
%
\chapter{Introduction and Motivation}
\index{Introduction@\emph{Introduction}}%

The most significant outcome of purely theoretical physics in the
last thirty years is string theory.  String theories, which
naturally live in $9+1$ dimensions, seem to provide a unified
perturbative description of all known interactions -- including
gravity.  In more recent years, however, it has been discovered
that even string theories appear to be unified into a mysterious
$10+1$ dimensional theory known as $\mathcal{M}$-theory, which is
believed to exist even at the non-perturbative level.

One of the most interesting and also most challenging aspects of
$\mathcal{M}$-theory is that it appears to admit a large number of
equivalent descriptions, each with its own set of degrees of
freedom and interactions.  Equivalences of this type are known as
\emph{dualities}.  By their very nature, dualities impose severe
difficulties on the formulation of a background independent
description of the theory, because some degrees of freedom are
more natural in certain backgrounds but none are universally
favored. Furthermore, $\mathcal{M}$-theory seems to incorporate a
radical concept called \emph{holography}, which requires that the
number of degrees of freedom in a region of space grows as the
area enclosing the region, instead of the volume. For this reason,
holography also obscures the background independence of the
theory, since one must choose a particular ``holographic screen"
on which to ``project" a description of the physics. Moreover, it
is not known the extent to which duality and holography interplay
in $\mathcal{M}$-theory.

Nonetheless, one conceivable way to formulate a theory which
embodies duality or holography in a background independent fashion
is to make manifest a symmetry whose action is equivalent to a
duality transformation or changing the choice of holographic
screen. In other words, one may choose to intentionally formulate
the theory in a highly redundant language. After all,
historically, this method has been proven correct in formulating
gauge theories, which are somewhat analogous. From this point of
view, different dual descriptions of the same physics or different
choices of holographic screen are merely different (partial) gauge
choices. Moreover, the reduction in degrees of freedom implied by
holography would presumably arise from eliminating all (or most
of) the gauge freedom. The problem is that duality and holography
imply such a rich structure that it is not known how to
parameterize any symmetry that could describe their full effect.
Although, there has been some evidence that suggests manifestly
duality invariant theories may be formulated in  backgrounds with
more than one time-like dimension \cite{ib:twotime}.

Following this line of reasoning, the key to discovering a
tractable background independent formulation of
$\mathcal{M}$-theory may be to study nontrivial extensions of
local gauge symmetry. Perhaps the most conservative step in this
direction is to study noncommutative gauge theories, which are
important in their own right because they are known to emerge from
string theory through various decoupling limits
\cite{sw:stri,sb:magn}. However, in the context of the present
discussion, noncommutative gauge theories could provide some clues
about the nature of whatever mysterious symmetry casts
$\mathcal{M}$-theory in its most symmetric form.

It turns out that noncommutative gauge theories do indeed contain
an essential ingredient which is necessary to accommodate
holography -- \emph{UV-IR mixing}. UV-IR mixing, or the UV-IR
connection, refers to a nondecoupling of UV and IR degrees of
freedom. Clearly, a nondecoupling of this type is required in any
holographic theory because the total number of degrees of freedom
in a region -- a quantity dependent on the UV -- is related to the
area enclosing the region -- a quantity dependent on the IR.  In
noncommutative gauge theories, UV-IR mixing arises from elementary
dipole degrees of freedom whose transverse length is proportional
to their center-of-mass momentum \cite{sb:magn}. Thus, UV dipoles
grow long in spatial extent and mediate instantaneous
long-distance interactions that are relevant in the IR. What is
more, the dipole character of noncommutative gauge theories is
intimately related to the structure imposed by gauge invariance.

However, while it may be plausible that the study of
noncommutative gauge invariance can shed some light on duality and
holography, the main objective of this work is to probe the
effects of UV-IR mixing in the context of noncommutative gauge
theories. In particular, we perturbatively analyze Wilsonian
integration in the UV regime defined by $\theta\Lambda^2\gg 1$.
Although not as fundamental in nature, this problem is quite
interesting and challenging due to the inherent involvement of
both spatial nonlocality and UV-IR nondecoupling. In fact, both of
these properties present major difficulties in the understanding
of noncommutative quantum field theories.

The most popular approach to studying noncommutative field
theories has been to work in the star product representation and
apply the conventional techniques that were originally developed
for local quantum field theories. Not surprisingly, UV-IR mixing
results in a tremendous amount of confusion regarding such things
as the renormalization of UV divergences
\cite{mr:uvdiv,cm:dual,cr:renorm}, the treatment of IR divergences
\cite{ms:nonp,mh:irdiv}, and Wilsonian integration
\cite{gp:wilsonrg,vv:wilson}. Moreover, the noncommutative gauge
invariance is not preserved order by order in the standard
perturbative expansion \cite{ls:gaug}. Rather, gauge invariance is
achieved by an infinite resummation of diagrams
\cite{ki:inte,ar:uvir,hl:trek}. Finally, there is a sort of
naturalness problem with the conventional approach in that the
intrinsic dipole structure of the elementary field quanta is not
completely clear, although some suggestive results have been
obtained \cite{ki:inte}.

On the other hand, in the matrix formulation of noncommutative
gauge theory, the noncommutative gauge invariance is manifest
\cite{mv:mean}, as is the dipole character of the elementary
quanta \cite{jn:dipo,en:renorm}. In fact, the matrix approach
allows for a clear separation between the quantum effects of UV
and IR dipoles. In particular, it has been shown that one can make
sense of Wilsonian integration despite UV-IR mixing, and
corrections to the quantum effective action resulting from
integrating out the UV states were explicitly calculated at both
the one- and two-loop orders. The resulting interactions were
found to dominate the long-distance behavior, which shed some
light on UV-IR mixing in noncommutative gauge theory, as well as
the nonanalytic dependence of the quantum theory on the
noncommutativity parameter $\theta$. Also, \cite{kk:bilocal}
provides a different point of view on how the matrix formulation
naturally leads to a bi-local representation. Thus, the dipole
structure imposed by the manifest gauge invariance of the matrix
formulation seems to hold the key to resolving the ambiguities
associated with UV-IR mixing.

In this work, we will review and build upon the recent
developments in the perturbative approach to Wilsonian integration
of noncommutative gauge theories in the matrix formulation. In
Chapter~\ref{sec:matrix}, we discuss the matrix formulation of
noncommutative gauge theories in some detail. The presentation is
based on \cite{rg:soli,ns:back}, although the proofs of
orthogonality and completeness of the Fourier matrix basis are
new. Then in Chapter~\ref{sec:pert}, which is essentially drawn
from \cite{jn:dipo,en:renorm}, we discuss the perturbative
Wilsonian treatment of noncommutative gauge theories in the UV
regime given by $\theta\Lambda^2\gg 1$. Finally,
Chapter~\ref{sec:renorm} addresses the divergence structure that
arises from the integration of UV modes. To some extent this part
is based on \cite{en:renorm}, although there are new discussions
concerning diagrammatical combinatorics and the relationship
between renormalization beyond one loop and UV-IR divergences. We
end in Chapter~\ref{sec:conc} with some discussion of our results
and outlook toward the open problem of understanding the IR regime
defined by $\theta\Lambda^2\lesssim 1$. The Appendices contain the
technical details and explicit calculations.

\chapter{Matrix Representation of Noncommutative Gauge
Theories} \label{sec:matrix}
\index{Instructions for Preparing Dissertations, Theses, and Reports%
@\emph{Instructions for Preparing Dissertations, Theses, and Reports}}%

\section{Introduction}

Before systematically developing the perturbative Wilsonian
treatment of noncommutative gauge theories in the matrix
formulation, we must first discuss the matrix representation of
the noncommutative plane. To that end, we will first define the
noncommutative plane algebraically and then briefly discuss its
novel properties.  We then derive the matrix representation of
functions defined on the noncommutative plane and discuss the
correspondence with the star product representation. Finally, we
formulate noncommutative gauge theories in the matrix
representation.

\section{Matrix Representation of the Noncommutative
Plane}\label{subsec:matrix}

The noncommutative $2p$-plane is most conveniently defined by its
algebra
\begin{equation}\label{eq:alg}
[x^i,x^j]=i\theta^{ij},
\end{equation}
where $x^i$ are the coordinates and $\theta^{ij}$ is an
antisymmetric tensor of $SO(2p)$.  Clearly, the novel feature of
the noncommutative plane is that the coordinates generally do not
commute. However, we may still think of the coordinates as any
other observable in our Hilbert space, and therefore, represent
them as infinite dimensional Hermitian matrices $\hat{x}^i$. The
difference in the noncommutative case is that Eq.~(\ref{eq:alg})
leads to a nontrivial uncertainty relation between the coordinate
operators
\begin{equation}\label{eq:unc}
\Delta x^i\Delta x^j\gtrsim\theta^{ij}.
\end{equation}
In words, Eq.~(\ref{eq:unc}) states that, on the noncommutative
plane, configurations that are localized in a given direction are
necessarily delocalized in the transverse direction.

Furthermore, when combined with the standard position-momentum
uncertainty relation, Eq.~(\ref{eq:unc}) implies that high
momentum configurations are necessarily to some extent
delocalized.  As we shall see in Chapter~\ref{sec:pert}, these
novel properties of the noncommutative plane will lead to spatial
nonlocality and UV-IR nondecoupling, which poses some technical
obstacles to understanding noncommutative quantum field theories.
However, the matrix representation is particularly well-suited to
describe this type of behavior, and will therefore be the
preferred language in which to formulate noncommutative field
theories.

In order to quantify the notion of ``configurations" that are
discussed above, we must first give a definition to arbitrary
functions of noncommutative coordinates. In the matrix
representation, we can think of functions  defined on the
noncommutative plane as matrices. Moreover, in analogy with the
familiar Fourier expansion of ordinary functions, it will prove
convenient to define functions of noncommutative coordinates by
means of an expansion in a special basis of matrices
\begin{equation}\label{eq:four}
F=\int\frac{{\rm
d}^{2p}k}{(2\pi)^{2p}}e^{ik\cdot\hat{x}}\widetilde{F}(k).
\end{equation}
When expressed this way, $F$ is a Weyl ordered function of
$\hat{x}^i$ since the exponential is.

The validity of Eq.~(\ref{eq:four}) follows from the orthogonality
and completeness relations that are derived in
Appendix~\ref{ap:orthocomp}
\begin{eqnarray}
(2\pi)^{p}({\rm det}\theta)^{1/2}{\rm Tr}\left(
e^{ik\cdot\hat{x}}e^{-ik^{\prime}\cdot\hat{x}}\right)&=&(2\pi)^{2p}
\delta^{2p}(k-k^{\prime})\label{eq:ortho}\\
\int\frac{{\rm d}^{2p}k}{(2\pi)^{2p}}
\left(e^{ik\cdot\hat{x}}\right)_{ij}\left(e^{-ik\cdot\hat{x}}\right)_{kl}
&=&\frac{\delta_{il}\delta_{kj}}{(2\pi)^{p}({\rm
det}\theta)^{1/2}}\label{eq:comp}.
\end{eqnarray}
Eqs.~(\ref{eq:ortho}) and (\ref{eq:comp}) imply that the Fourier
coefficients are given by
\begin{equation}
\widetilde{F}(k)=(2\pi)^{p}({\rm det}\theta)^{1/2}{\rm Tr}\left(
e^{-ik\cdot\hat{x}}F\right).
\end{equation}
In what is to follow, we shall \emph{define} $f(\hat{x})$ as the
matrix whose Fourier coefficients are
\begin{equation}\label{eq:coef}
\tilde{f}(k)=\int {\rm d}^{2p}x\,e^{-ik\cdot x}f(x).
\end{equation}
Eq.~(\ref{eq:coef}) is the canonical choice because it implies
that, as a function of its argument, $f(\hat{x})$ reduces to
$f(x)$ in the commutative limit. However, perhaps the most
convenient aspect of Eq.~(\ref{eq:coef}) is that it implies
\begin{equation}
(2\pi)^{p}({\rm det}\theta)^{1/2}{\rm Tr}f(\hat{x})=\int {\rm
d}^{2p}x\, f(x),
\end{equation}
which will play a crucial role in constructing the action for
noncommutative field theories.

An immensely important consequence of Eq.~(\ref{eq:coef}) is that
it immediately provides a one-to-one correspondence between matrix
functions and ordinary functions
\begin{equation}\label{eq:co}
f(\hat{x})\longleftrightarrow f(x).
\end{equation}
Thus, to $f(\hat{x})$, we may attribute a profile in position and
momentum space given by $f(x)$ and $\tilde{f}(k)$, respectively.
It is, however, important to realize that there is a crucial
difference between the algebra of functions defined on the
noncommutative plane and those defined on the ordinary plane. For
example, consider the product of two matrix functions $f(\hat{x})$
and $g(\hat{x})$. From Eqs.~(\ref{eq:coef}) and (\ref{eq:co}) it
follows that
\begin{equation}\label{eq:prod}
f(\hat{x})g(\hat{x})\longleftrightarrow (f\star g)(x)\equiv
\exp\left(\frac{i}{2}\partial_i\theta^{ij}\partial_j^\prime\right)f(x)g(x^\prime)\bigg|_{x^\prime=x}.
\end{equation}
Thus, if we choose to represent the matrix functions $f(\hat{x})$
and $g(\hat{x})$ by the ordinary functions $f(x)$ and $g(x)$, we
must deform the ordinary product into the star product as shown in
Eq.~(\ref{eq:prod}). This choice corresponds to the star product
representation. From the perspective of the Weyl ordering
convention, Eq.~(\ref{eq:prod}) states that the Weyl ordered
product of two Weyl ordered functions $f(\hat{x})$ and
$g(\hat{x})$ is $(f\star g)(\hat{x})$. More physically,
Eq.~(\ref{eq:prod}) represents the quantum nature of the
noncommutative plane.

Although there is a complete duality between the matrix and star
product representations, it is generally more convenient to work
in the matrix representation in order to avoid the technical
complications that are introduced by the star product.
Furthermore, there is an added benefit to the matrix formulation
in that the noncommutative gauge invariance, which is discussed in
Section~\ref{subsec:form}, is manifest order by order in
perturbation theory. Therefore, from this point on, we will work
in the matrix representation.

\section{Matrix Formulation of Noncommutative Gauge
Theories}\label{subsec:form}

In Section~\ref{subsec:matrix} we introduced all of the
ingredients necessary to construct noncommutative gauge theories
in the matrix formulation. The correspondence given by
Eq.~(\ref{eq:co}) guarantees that an arbitrary gauge field
configuration can be represented by
\begin{equation}\label{eq:field}
A_\mu(\hat{x},t)=\int\frac{{\rm d}^{2p}k}{(2\pi)^{2p}}
e^{ik\cdot\hat{x}}\otimes\widetilde{A}_\mu(k,t).
\end{equation}
The tensor product in Eq.~(\ref{eq:field}) refers to the product
between the matrix structure of the exponential function of
$\hat{x}^i$ and the matrix structure of the Fourier coefficients,
which we take to live in the adjoint representation of $U(N)$.

When formulating noncommutative gauge theories in the matrix
representation, it is convenient to define the quantity
\begin{equation}\label{eq:defx}
X^i(t)=\hat{x}^{i}\otimes{\rlap{1} \hskip 1.6pt
\hbox{1}}_N+\theta^{ij}A_j(\hat{x},t).
\end{equation}
Eq.~(\ref{eq:defx}) implies that gauge transformations are
properly realized if and only if $X^i$ transforms in the adjoint
representation of the noncommutative gauge group, which we take to
be $U(N)_{\rm NC}$
\begin{equation}
X^i\rightarrow U^\dagger X^iU\iff A_j\rightarrow U^\dagger
A_jU+iU^\dagger\partial_jU.
\end{equation}
Furthermore, space derivatives are simply given by taking the
commutator
\begin{equation}\label{eq:cov}
[X^i(t),\Phi(\hat{x},t)]=i\theta^{ij}(D_j\Phi)(\hat{x},t),
\end{equation}
where $\Phi(\hat{x},t)$ can be any field transforming in the
adjoint representation of the gauge group and
$D_j\Phi\equiv\partial_j\Phi-iA_j\star\Phi+i\Phi\star A_j$ is the
gauge covariant derivative of $\Phi$. Of course, time derivatives
are defined in the standard way, and gauge covariance is achieved
by introducing a gauge field $A_0(t)=A_0(\hat{x},t)$ transforming
as
\begin{equation}
A_0\rightarrow U^\dagger A_0U+iU^\dagger\dot{U}
\end{equation}
under noncommutative gauge transformations.

Using the definitions of the fields $X^i$ and $A_0$, it is
straight forward to show that
\begin{eqnarray}\label{eq:str}
\dot{X^i}-i\left[A_0,X^i\right] & = & \theta^{ij}F_{0j}(\hat{x})\nonumber\\
-i\left[X^i,X^j\right] & = & \theta^{ik}\theta^{jl}
\left(F_{kl}(\hat{x})-\theta^{-1}_{kl}\right),
\end{eqnarray}
where $F_{\mu\nu} \equiv
 \partial_\mu A_\nu - \partial_\nu A_\mu - iA_\mu\star A_\nu +iA_\nu\star A_\mu$ is the noncommutative gauge field
 strength.  Therefore, up to total derivative terms, the Lagrangian for $2p+1$ dimensional noncommutative Yang-Mills (NCYM)
\begin{eqnarray}\label{eq:lag}
 L = \int {\rm d}^{2p}x\,{\rm tr}_N\left(\frac{1}{2}G^{ij}F_{0i}(x)F_{0j}(x)-\frac{1}{4}G^{ij}G^{kl}F_{ik}(x)F_{jl}(x)+\textrm{total
 deriv}\right),\nonumber\\
\end{eqnarray}
can be recast in terms of a $0+1$ dimensional theory of matrix
fields
\begin{eqnarray}\label{eq:mat}
 L={\rm
Tr}\left(\frac{1}{2}\left(\dot{X^i}-i\left[A_0,X^i\right]\right)^{2}+\frac{1}{4}\left[X^i,X^j\right]\left[X^i,X^j\right]\right),
\end{eqnarray}
where the inverse spatial metric of the field theory
$G^{ij}=\theta^{ik}\theta^{kj}$. Note that for convenience, we
have chosen units such that $(2\pi)^{2p}{\rm det}(\theta)=1$.

At this point, some remarks are in order.  Eq.~(\ref{eq:lag})
shows that the matrix description of NCYM given by
Eq.~(\ref{eq:mat}) includes some contributions that are total
space derivatives from the field theory perspective. While this is
expected to be significant non-perturbatively, at least at the
perturbative level, it is irrelevant because the equations of
motion are unchanged. Thus, we can capture the perturbative
behavior of NCYM theory by studying Eq.~(\ref{eq:mat}), which is
exactly what we shall do.  Another point is that this same
construction can be generalized in an obvious manner to include
fermions or any other matter field transforming in the adjoint
representation of the gauge group. For example, the theory can be
easily supersymmetrized, which we will explore in more detail in
Chapters~\ref{sec:pert} and \ref{sec:renorm}. Lastly, it should be
noted that we can describe theories with $d$ additional
commutative spatial dimensions by simply promoting
Eq.~(\ref{eq:lag}) from $0+1$ dimensions to $d+1$ dimensions. We
will frequently rely on this correspondence in
Chapters~\ref{sec:pert} and \ref{sec:renorm} when we discuss $3+1$
dimensional noncommutative theories.

Before moving on to the perturbative behavior of NCYM, it is worth
while discussing the noncommutative gauge invariance in more
detail.  From the tensor product structure of Eq.~(\ref{eq:defx}),
we see that the noncommutative gauge group, which we take to be
$U(N)_{\rm NC}$, can be decomposed as $U(\infty)\otimes U(N)$,
both factors generally being time-dependent. The $U(N)$ factor
encodes the spatially constant part of the gauge transformation,
while the $U(\infty)\equiv U(1)_{\rm NC}$ factor encodes the
spatial dependence of the gauge transformation. However, due to
the matrix structure of $U(\infty)$, which connects spatial points
nonlocally, it is clear that noncommutative groups are a much
larger group of symmetries than ordinary groups. As such, the
structure imposed by the noncommutative gauge invariance will be
much more rigid than that of ordinary gauge theories, and in fact
all orders of the conventional perturbative expansion will
contribute to gauge invariant quantities. Understanding this
structure will be a central theme of this work.

\chapter{Perturbative Wilsonian Formalism for Noncommutative Gauge
Theories} \label{sec:pert}
\index{How to Use the utdiss2 Package%
@\emph{How to Use the utdiss2 Package}}%

\section{Introduction}

In Chapter~\ref{sec:matrix}, we showed how noncommutative gauge
theories are formulated in the matrix representation.  We will now
develop Wilsonian perturbation theory in this language in order to
probe the physical effects of UV-IR mixing. We begin by rewriting
Eq.~(\ref{eq:mat}) in the background field gauge, which is
generally convenient in the Wilsonian approach. We then derive the
propagator for the UV modes and utilize it to calculate both the
leading and the next to leading order one- and two-loop
corrections to the Wilsonian quantum effective action. Finally, we
determine the general gauge invariant structure of perturbation
theory that is valid to all loop orders in the UV regime defined
by $\theta\Lambda^2\gg 1$.

\section{Background Field Gauge Fixing}

Ultimately, we are interested in performing a Wilsonian
integration of the UV modes. In order to separate the quantum
effects of UV and IR states, it is convenient to work in the
background field language.  In this approach, we expand the fields
$A_0=B_0+A$ and $X^{i}=B^{i}+Y^{i}$, where the background fields
$B_0$ and $B^{i}$ contain the IR degrees of freedom, while the
fluctuating fields $A$ and $Y^{i}$ contain the UV degrees of
freedom. Of course, we can think of the background fields in the
usual way, $B_0=A_0(\hat{x})$ and $B^i=\hat{x}^{i}\otimes{\rlap{1}
\hskip 1.6pt \hbox{1}}_N+\theta^{ij}A_j(\hat{x})$, using the
definition given by Eq.~(\ref{eq:field}). The fluctuating fields
then have the natural interpretation as fluctuations in $A_0$ and
$A_i$. The imposition of the Wilsonian cutoff is somewhat subtle,
however, so that point will be deferred until
Section~\ref{subsec:prop}.

In order to define the functional integral over $A$ and $Y^{i}$,
we must gauge fix the Lagrangian. This can be accomplished by
adding both a gauge fixing and the corresponding ghost term to the
Lagrangian
\begin{eqnarray}
L_\mathrm{gf}+L_\mathrm{gh} & = & {\rm
Tr}\left(-\frac{1}{2}\left(-\dot{A}-i\left[B^i,Y^i\right]\right)^2
+ \dot{\bar{c}}\Big(\dot{c}-i[A,c]\Big)
+\left[B^i,\bar{c}\right]\left[X^i,c\right]\right).\nonumber\\
\end{eqnarray}
Note that, for convenience, we have used the residual gauge
symmetry of the background fields to set $B_0=0$, although we may
restore $B_0$ at any time simply by gauge covariantizing the time
derivatives. However, it is important to realize that no
additional gauge fixing terms are required because the background
fields are not to be integrated out. Upon expanding in
fluctuations, the action takes the form $L=L_0+L_1+L_2+L_3+L_4$
where
\begin{eqnarray}\label{eq:ord}
L_0 & = & {\rm
Tr}\left(\frac{1}{2}{\dot{B}}^{i2}+\frac{1}{4}[B^{i},B^{j}][B^{i},B^{j}]\right);
\nonumber\\
L_2 & = & {\rm Tr}\bigg(\frac{1}{2}{\dot{Y}}^{j2}+\frac{1}{2}[B^i,Y^j]^{2}-\frac{1}{2}{\dot{A}}^{2}-\frac{1}{2}[B^i,A]^{2}-2i{\dot{B}}^i[A,Y^i]\nonumber\\
& &  \qquad +[B^i,B^j][Y^i,Y^j]+\dot{\bar{c}}\dot{c}+[B^i,\bar{c}][B^i,c] \bigg); \nonumber\\
L_3 & = & {\rm Tr}\bigg([B^i,Y^j][Y^i,Y^j]-[B^i,A][Y^i,A]-i{\dot{Y}}^i[A,Y^i]\nonumber\\
& &  \qquad -i\dot{\bar{c}}[A,c]+[B^i,\bar{c}][Y^i,c] \bigg); \nonumber\\
L_4 & = & {\rm
Tr}\left(\frac{1}{4}[Y^i,Y^j][Y^i,Y^j]-\frac{1}{2}[A,Y^i]^{2}\right).
\end{eqnarray}

Notice that we have neglected the linear term $L_1$, which
generically will contribute to the dynamics of the background
field through tadpoles. In the language of perturbation theory,
this amounts to corrections that are both higher order in the
gauge coupling and higher order in derivatives of the background
field. Therefore, the linear term, is suppressed if we require
both that the coupling be sufficiently weak so that the loop
corrections are small and that the states we integrate out be of
sufficiently high momenta relative to the scale set by the
background so that the higher derivative corrections are small, as
well. In fact, both of these conditions will be met in our
perturbative Wilsonian approach, so for the sake of simplicity, we
have neglected the linear interactions in Eq.~(\ref{eq:ord}). Of
course, consistency requires $L_1$ be included in a complete
perturbative analysis, but it does not contribute to any of the
quantities that we calculate in this work.

The structure of the quadratic term $L_2$ plays a central role in
perturbation theory. We see from Eq.~(\ref{eq:ord}) that all of
the fluctuating fields have similar quadratic terms up to
interactions proportional to $i{\dot{B}}^{i}$ and $[B^{i},B^{j}]$,
which Eq.~(\ref{eq:str}) implies are proportional to the
background gauge field strength. As is well known, the background
field dependence of the terms quadratic in the fluctuating fields
can either be treated exactly or perturbatively, depending on the
definition of the propagator. In our calculation, it will be most
convenient to treat the field strength terms perturbatively, while
absorbing the remaining background dependence into the propagator.
From a physical standpoint, our treatment of $L_1$ and $L_2$
corresponds to a derivative expansion of the background field,
which is valid when UV modes are integrated out. In fact, in this
work, we only consider the effects of integrating out states of
energy and momentum much higher than any other scale in the
problem, including the noncommutativity parameter $\theta$.

Finally, before deriving the propagator for the fluctuating
fields, let us discuss the structure of expectation values in the
background field language.  When we compute the effects of the
interaction terms in Eq.~(\ref{eq:ord}) perturbatively, we will
encounter expectation values involving both background and
fluctuating fields.  Since the background fields are still quantum
operators in the Wilsonian scheme, we cannot simply remove them
from the expectation value.  However, since the background and
fluctuating fields contain entirely different degrees of freedom,
the expectation value factorizes into a product of the expectation
value of the background fields and the expectation value of the
fluctuating fields.  In order to integrate out only the
fluctuating fields, we evaluate the expectation value of the
fluctuating fields and interpret the residual expectation value of
background fields as arising from a term in the quantum effective
action.

\section{Propagator in the Matrix
Representation}\label{subsec:prop}

As discussed above, all of the fluctuating fields,
$\Phi=(Y^i,A,\bar{c},c)$ corresponding to gauge field degrees of
freedom, appear similarly in quadratic terms of the form
\begin{equation}\label{eq:adj}
{\rm
Tr}\left(\frac{1}{2}\dot{\Phi}^2+\frac{1}{2}\left[B^i,\Phi\right]^2\right).
\end{equation}
In Eq.~(\ref{eq:adj}), $\Phi$ is manifestly in the adjoint
representation of the noncommutative gauge group. However, it will
prove convenient to express the adjoint representation as the
tensor product of the fundamental and antifundamental
representations. In index notation, the tensor product realization
of Eq.~(\ref{eq:adj}) becomes
\begin{eqnarray}
 & & \frac{1}{2}\Phi_b^a\left(-\delta_c^b\delta_a^d\frac{{\rm d}^2}{{\rm d}t^2}-B^{ib}_eB^{ie}_c\delta_a^d-\delta_c^bB^{id}_eB^{ie}_a+2B^{ib}_cB^{id}_a\right)\Phi_d^c,
 \end{eqnarray}
 which implies the following matrix structure
\begin{equation}
 \frac{1}{2}\Phi^T\left(-{\rlap{1} \hskip 1.6pt \hbox{1}}\otimes {\rlap{1} \hskip 1.6pt \hbox{1}} \frac{{\rm d}^2}{{\rm d}t^2}-\left(B^i\otimes {\rlap{1} \hskip 1.6pt \hbox{1}}-{\rlap{1} \hskip 1.6pt \hbox{1}}\otimes B^i\right)^2\right)
\Phi.
\end{equation}

In the Wilsonian scheme, we are only  interested in integrating
out virtual states with frequencies $\omega\gg 1/T$, $T$ being the
time scale set by the background. For these high frequency modes,
the backreaction coming from the background time dependence is a
subleading effect. Therefore, the matrix propagator for virtual
states with frequencies above a Wilsonian cutoff, $\Lambda\gg
1/T$, can be expressed in the following Fourier integral form
\begin{equation}
G(t-t^\prime)=\int_\Lambda\frac{{\rm
d}\omega}{2\pi}\frac{e^{-i\omega
(t-t^\prime)}}{\omega^2-M^2}+\ldots
\end{equation}
where  $M^2=\left(B^i\otimes {\rlap{1} \hskip 1.6pt
\hbox{1}}-{\rlap{1} \hskip 1.6pt \hbox{1}}\otimes B^i\right)^2$
and the $\ldots$ represent subleading terms that are suppressed by
factors of $(T\Lambda)^{-1}\ll 1$.  In the following, we consider
only the leading order term.

To relate this $0+1$ dimensional matrix quantity to a $2p+1$
dimensional field theory quantity, we choose a convenient
representation which is derived in Appendix~\ref{ap:prop}
{\setlength\arraycolsep{2pt}
\begin{eqnarray}\label{eq:rep}
& & \frac{1}{\omega^2-M^2} \nonumber\\
& = & \int\frac{{\rm d}^{2p}k}{(2\pi)^{2p}}e^{-ik\cdot(B\otimes
1-1\otimes B)}
\int_{\theta\Lambda}{\rm d}^{2p}x \frac{e^{ik\cdot (x-x^\prime)}}{\omega^2-(x-x^\prime)^2}\\
& = & \int\frac{{\rm d}^{2p}k}{(2\pi)^{2p}}e^{-ik\cdot B}\otimes
e^{ik\cdot B}\int_{\theta\Lambda} {\rm d}^{2p}x\,e^{ik\cdot
(x-x^\prime)}\widetilde{G}\left(\omega,\theta^{-1}(x-x^\prime)\right),\nonumber
\end{eqnarray}}\\
where $\widetilde{G}(\omega,p)=(\omega^2-p_i G^{ij} p_j)^{-1}$ is
the field theory momentum space propagator for a massless state.
As discussed in Appendix~\ref{ap:prop}, there is a lower cutoff
applied to the integral over $x$ such that
$|x-x^\prime|>\theta\Lambda\gg{\theta}/{L}$ where $L$ is the
length scale set by the curvature of the background. Moreover, the
physical reason for imposing the cutoff on the position integral
is now clear, since position apparently plays the role of
momentum. Putting everything together, the matrix propagator can
be written in the following form
\begin{eqnarray} \label{eq:prop}
\lefteqn{ G(t-t^\prime)=\int_{\theta\Lambda} d^{2p}x
\int_\Lambda\frac{d\omega}{2\pi}\int\frac{d^{2p}k}{(2\pi)^{2p}}
e^{-i\omega(t-t^\prime)+
ik\cdot (x-x^\prime)} } \nonumber\\
 & & \qquad\qquad\qquad {}\times \widetilde{G}\left(\omega,\theta^{-1}(x-x^\prime)\right)e^{-ik\cdot B(t)}\otimes e^{ik\cdot B(t)}.\qquad
\end{eqnarray}

We can now identify the various ingredients of the $0+1$
dimensional propagator given by Eq.~(\ref{eq:prop}) from a $2p+1$
dimensional perspective. As suggested by the notation,
$(\omega,k)$ is to be identified with the spacetime
energy-momentum, and likewise, $(t,x)$ is the corresponding
spacetime coordinate. The integral over $x$ is then understood in
terms of the nonlocality of the noncommutative field theory, but
perhaps more surprising is the role played by the field theory
propagator, $\widetilde{G}$. Evidently, the small-$k$ -- large-$x$
region of the integral corresponding to low-momentum --
large-distance receives contributions from \emph{high} momentum
field theory states and vice-versa. Actually, this type of
behavior has a very natural interpretation in terms of the dipole
degrees of freedom that we expect from the decoupling limit in
which noncommutative gauge theory emerges from string theory.

In the decoupling limit, the noncommutative field quanta can be
thought of as dipoles with a transverse size proportional to the
center of mass momentum $(x-x^\prime)^i=\theta^{ij}p_j$. It is
clear that this effect is encoded in Eq.~(\ref{eq:prop}), since
the momentum argument of the field theory propagator
$\widetilde{G}(\omega,p)$ is $p=\theta^{-1}(x-x^\prime)$. It is
also clear that, due to the Fourier integral over position, these
dipole states probe a transverse momentum scale $k_i\sim
1/(x-x^\prime)^i$. Combining the two relations, we arrive at
$1\sim p_i\theta^{ij}k_j$, which is equivalent to the stationarity
of the Moyal phase factors that appear in the star product
formulation \cite{ms:nonp}. In essence, this relation means that
integrating out high momentum states can lead to low momentum
effects -- a symptom of UV-IR mixing. Thus, it seems that
Eq.~(\ref{eq:prop}) naturally describes the dipole degrees of
freedom that appear in noncommutative gauge theories.

However, it is important to realize that the representation of the
propagator given by Eq.~(\ref{eq:prop}) is only valid for dipoles
of high energy and momentum. More precisely, if the background
changes on time and length scales $T$ and $L$, respectively, we
can only integrate over frequencies $\omega \gg 1/T$ and momenta
$p=\theta^{-1}(x-x^\prime) \gg 1/L$; otherwise, the time
derivatives and commutators involving the background field that
were dropped in the derivation of Eq.~(\ref{eq:prop}) are no
longer negligible. Therefore, the cutoff $\Lambda$ is chosen such
that $\Lambda\gg 1/T \,\textrm{and}\, 1/L$, in which case the
higher order commutator and time derivative corrections are
suppressed by factors of $(L\Lambda)^{-1}\,\,\textrm{and}\,\,
(T\Lambda)^{-1}\ll 1$. Moreover, since the cutoff is chosen
relative to the scale of the background, $\Lambda$ is naturally
interpreted in the Wilsonian sense.

The matrix structure of the $0+1$ dimensional propagator, which is
contained entirely in the tensor product of operators of the form
$\exp(ik\cdot B)$, also has an important field theory
interpretation. As derived in Appendix~\ref{ap:line}, we can
identify
\begin{eqnarray}\label{eq:line}
 e^{ik\cdot B(t)}=e^{ik\cdot\hat{x}\otimes 1_N+ik\cdot\theta\cdot
A(\hat{x},t)}\longleftrightarrow P_\star e^{i\int_0^{1}{\rm
d}\sigma k\cdot\theta\cdot A(x+\sigma\theta\cdot k,t)}\star
e^{ik\cdot
 x},
\end{eqnarray}
where $P_\star$ denotes path ordering of the exponential using the
star product.  This object transforms in the adjoint under gauge
transformation, and in particular, the trace is gauge invariant
\begin{equation}\label{eq:rho1}
{\rm Tr}\left(e^{ik\cdot B(t)}\right)\longleftrightarrow \int {\rm
d}^{2p}x e^{ik\cdot x} {\rm tr}_N\left(P_\star e^{i\int_0^{1}{\rm
d}\sigma k\cdot\theta\cdot A(x+\sigma\theta\cdot k,t)}\right).
\end{equation}\\
We immediately recognize this object as the Fourier transform of
an open Wilson line . In fact, this structure was essentially
guaranteed by the noncommutative gauge invariance \cite{ni:obse}.

When we use the matrix propagator for perturbative calculations of
the quantum effective action, we will frequently encounter the
Fourier transform of Eq.~(\ref{eq:rho1}). Following
\cite{mv:mean}, we define the operator
\begin{equation}\label{eq:rho2}
\rho(x,t)=\int\frac{{\rm d}^{2p}k}{(2\pi)^{2p}}e^{ik\cdot x}{\rm
Tr}\left(e^{-ik\cdot B(t)}\right).
\end{equation}
Although $\rho(x)$ is generally a spatially nonlocal field theory
operator, for $\theta\cdot k$ sufficiently small such that
Eq.~(\ref{eq:prop}) is valid, it is approximately local on length
scales given by the background configuration, as can be easily
seen from Eqs.~(\ref{eq:rho1}) and (\ref{eq:rho2}). In fact, all
gauge invariant Wilson line operators, which differ only by extra
operator insertions, will share this property. For example, an
insertion of an arbitrary operator $\mathcal{O}$ into the end of
the Wilson line gives
\begin{equation}\label{eq:insertion}
{\rho}_\mathcal{O}(x,t)=\int\frac{{\rm
d}^{2p}k}{(2\pi)^{2p}}e^{ik\cdot x}{\rm
Tr}\left(\mathcal{O}(t)e^{-ik\cdot B(t)}\right).
\end{equation}

The interpretation of the matrix propagator in terms of dipole
degrees of freedom is made more concrete in
Sections~\ref{subsec:one} and \ref{subsec:two} by calculating the
leading and next to leading order one- and two-loop corrections to
the Wilsonian quantum effective action. We will find that
integrating out UV virtual states gives rise to long-distance
interaction terms that are naturally interpreted in the dipole
context discussed above.

\section{One-loop Corrections to the Effective
Action}\label{subsec:one}

\begin{figure}
\centering \subfigure[Single matrix diagram is manifestly gauge
invariant and implicitly contains all leading background
dependence.]{\label{fig:subfig:a}\begin{minipage}[b]{.45\textwidth}\centering
\includegraphics[scale=0.25]{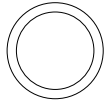}\end{minipage}}
\hspace{1cm} \subfigure[Gauge invariance achieved by summing over
all background insertions on both the outer and inner
boundaries.]{\label{fig:subfig:b}\begin{minipage}[b]{.45\textwidth}\centering
\includegraphics[scale=0.25]{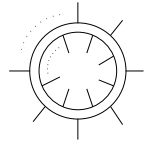}\end{minipage}}
\caption{One-loop contributions in the matrix versus the star
product approach.}
\end{figure}

We begin the computation of the quantum effective action at
one-loop, where the advantage of the matrix formulation becomes
immediately clear. The leading one-loop contribution is manifestly
gauge invariant and can be expressed in a single diagram drawn in
't~Hooft double line notation as shown in Fig.~\ref{fig:subfig:a}.
This is to be contrasted with the field theory star product
approach in which an infinite number of diagrams of the form shown
in Fig.~\ref{fig:subfig:b} must be summed up in order to achieve
gauge invariance \cite{ki:inte,ar:uvir,hl:trek}.

Using the matrix representation of the propagator given by
Eq.~(\ref{eq:prop}), the evaluation of the matrix diagram is
simple. The contraction of matrix indices, as indicated by the
double line diagram, gives a double trace contribution
proportional to {\setlength\arraycolsep{2pt}
\begin{eqnarray}\label{eq:trac}
 & & (N_B-N_F)\int {\rm d}t {\rm d}t^\prime\delta(t-t^\prime){\rm Tr}\log G(t-t^\prime) \\
 & = & (N_B-N_F)\int {\rm d}t {\rm d}^{2p}x_1 {\rm d}^{2p}x_2 \rho(x_1,t)\rho(x_2,t)\int\frac{{\rm
 d}\omega}{2\pi}\log\widetilde{G}(\omega,\theta^{-1}x_{12}),
                   \nonumber
\end{eqnarray}}\\
where $x_{12}\equiv x_1-x_2$, and $N_B$ and $N_F$ are the numbers
of on-shell bosonic and fermionic polarization states,
respectively. Note that, although we have discussed only the gauge
degrees of freedom explicitly, fermions and other matter fields
will generally contribute quantities of the same form as the pure
gauge calculation, but they will differ in the constant of
proportionality, as in Eq.~(\ref{eq:trac}) above.  Furthermore,
the integrals are always assumed to be cutoff as previously
discussed, even though the cutoff will often be suppressed.

Now let us understand the structure of Eq.~(\ref{eq:trac}) a bit
more in terms of the conventional field theory diagrams that are
illustrated in Fig.~\ref{fig:subfig:b}. First of all, we choose to
expand $\rho(x)={\rm tr}_{N}({\rlap{1} \hskip 1.6pt
\hbox{1}})+\Delta(x)$. The significance of $\Delta(x)$ is that it
contains only fluctuations of the background field around the
vacuum state. In particular, $\Delta(x)$ vanishes for trivial
configurations gauge equivalent to $A_{i}(x)=0$, which can be seen
from Eqs.~(\ref{eq:rho1}) and (\ref{eq:rho2}). Therefore, the
field theory interpretation of $\Delta(x)$ is that it represents
the gauge invariant contribution from the insertions of the
background gauge field into a boundary of the loop. On the other
hand, the constant term of $\rho(x)$ is gauge field independent,
and therefore, must descend from field theory diagrams with no
background insertions into the corresponding outer or inner
boundary.

For example, we can conclude that the ${\Delta}^{0}$ term involves
no insertions on either the outer or the inner boundary, and
therefore, comes from field theory vacuum diagrams. Using the same
reasoning, we find that the ${\Delta}^{1}$ terms involve
background insertions on only one boundary, and therefore, are due
to non-vacuum planar field theory diagrams. Finally, the
${\Delta}^{2}$ term involves insertions on both the outer and the
inner boundary, and therefore, arises from nonplanar field theory
diagrams. Thus, the single matrix diagram in
Fig.~\ref{fig:subfig:a} contains contributions from both planar
and nonplanar field theory diagrams.

However, Eq.~(\ref{eq:trac}) only reproduces the leading order
terms of the expansion in external momenta, as can be verified by
a direct field theory calculation \cite{mv:mean}. The reason is
that in deriving Eq.~(\ref{eq:prop}), the matrix formulation
naturally leads to an expansion in commutators and time
derivatives. The subleading terms, as we have seen,  are
suppressed by factors of $(L\Lambda)^{-1}$, $1/L$ being the scale
of the external momenta and $\Lambda$ the scale of the Wilsonian
cutoff. This must correspond to the expansion in external momenta
of the field theory propagators in the loop diagrams, since the
expansion parameter is the same. On the other hand, the external
momentum dependence of the Moyal phase factors is determined
exactly by noncommutative gauge invariance. In terms of the
structure that we discussed in Section~\ref{subsec:prop}, the
expansion in the external momenta of the propagators manifests
itself as higher dimensional operator insertions into the end of
the Wilson lines as in Eq.~(\ref{eq:insertion}), while the Wilson
lines themselves are the manifestation of both the additional
propagators and the Moyal phase factors associated to the external
insertions. Thus, as alluded to earlier, the physical nature of
our approximation is that of a standard derivative expansion of
the background. In fact, order by order, the matrix approach
reproduces the momentum expansion of the field theory propagators
if higher order commutators and time derivatives are retained.

Back to the evaluation of Eq.~(\ref{eq:trac}), the vacuum
diagrams, corresponding to the $\Delta^0$ term, are divergent. It
is easy to see that they are proportional to
\begin{eqnarray}\label{eq:vacdiv}
& & (N_B-N_F)N^2V\int{\rm d}t\frac{{\rm d}\omega}{2\pi} {\rm
d}^{2p}x_{12}
\log{\widetilde{G}(\omega,{\theta}^{-1}x_{12})}\nonumber\\
&=& (N_B-N_F)N^2V\int{\rm d}t\frac{{\rm d}\omega {\rm
d}^{2p}p}{2\pi(2\pi)^{2p}}\log{\widetilde{G}(\omega,p)}.
\end{eqnarray}
In fact, this is nothing but the usual one-loop UV vacuum
divergence that is familiar from ordinary field theories. The
leading one-loop divergent contribution from other non-vacuum
planar diagrams, which corresponds to the $\Delta^1$ term,
vanishes because $\int {\rm d}^{2p}x\Delta(x)=0$, as can be seen
from Eq.~(\ref{eq:rho2}). Thus, in the case of planar field theory
diagrams, the leading one-loop correction reduces to known
results.

\begin{figure}
\centering \psfrag{x}{$x_1$} \psfrag{y}{$x_2$}
\includegraphics[height=1cm]{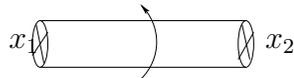}
\caption{High momentum virtual dipoles grow long in the transverse
direction and mediate instantaneous interactions between distant
background fluctuations at $x_1$ and $x_2$.} \label{fig:cyli}
\end{figure}

On the other hand, the nonplanar diagrams represented by the
${\Delta}^{2}$ interaction
\begin{equation}\label{eq:non1}
(N_B-N_F)\int{\rm d}t{\rm d}^{2p}x_1{\rm d}^{2p}x_2
\Delta(x_1,t)\Delta(x_2,t)\int\frac{{\rm
d}\omega}{2\pi}\log{\widetilde{G}(\omega,{\theta}^{-1}x_{12})},
\end{equation}
demonstrates an entirely new effect. This term illustrates how UV
dipoles can mediate long-distance interactions. When the virtual
dipoles in the loop have high momentum, Fig.~\ref{fig:subfig:a}
``stretches out'' into a long cylinder that joins distant points
$x_1$ and $x_2$. Each boundary of the cylinder contributes a trace
which yields a gauge invariant Wilson line operator corresponding
to low momentum background insertions into the field theory
diagrams. This process is depicted in Fig.~\ref{fig:cyli}.
\emph{Thus, we can interpret the double lines of the matrix
diagram as literally representing the end points of the virtual
dipole quanta as they propagate around the loops}.

In fact, by performing the integration over frequency, we can
calculate the interaction strength between $\Delta(x_1)$ and
$\Delta(x_2)$
\begin{equation}
\int \frac{{\rm
d}\omega}{2\pi}\log\widetilde{G}(\omega,\theta^{-1}x_{12})\sim|x_1-x_2|+\textrm{constant}.
\end{equation}
Thus, in the presence of this term, the theory is strongly
interacting at long distances.  This fact has been recognized in
\cite{mv:mean}, and it was shown that these strong long-distance
interactions are due to the leading IR pole singularities that
appear in nonsupersymmetric noncommutative theories. The
appearance of nonanalytic behavior in external momenta has also
been discussed in the star product formulation in
\cite{mr:uvdiv,cm:dual,cr:renorm,ms:nonp,mh:irdiv,gp:wilsonrg,vv:wilson,ls:gaug,ki:inte,ar:uvir,hl:trek}.
However, if the theory is supersymmetric, then $N_B=N_F$ and the
leading order one-loop interaction given by Eq.~(\ref{eq:trac})
vanishes.

We must now consider the next to leading order one-loop
contribution, which has also been discussed from the star product
perspective in \cite{ar:uvir}. As alluded to earlier, the precise
result requires that we keep the next to leading order commutators
and time derivatives that were dropped in the derivation of the
propagator, as well as extra insertions of the background field
strength coming from terms in $L_2$ that were also excluded from
the propagator. However, power counting as well as symmetry
arguments imply that the next to leading order one-loop
contribution will be of the same order as Fig.~\ref{fig:subfig:a}
with two extra insertions of the field strength
\begin{center}
\includegraphics[scale=0.25]{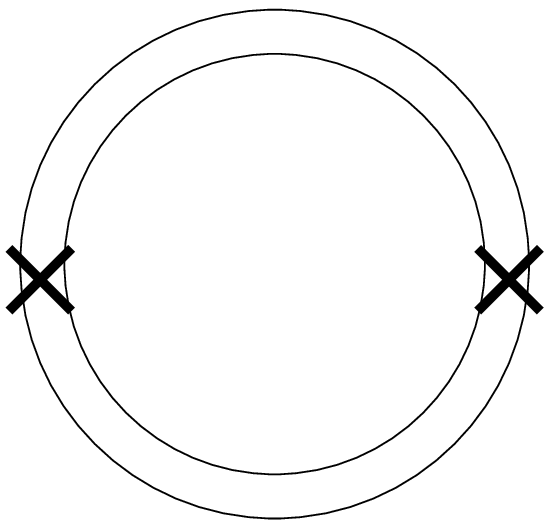}.
\end{center}
This contribution alone suffices to demonstrate the qualitative
features of the next to leading order one-loop behavior. It has
the added virtue that the calculation can be done with the leading
order propagator Eq.~(\ref{eq:prop}) because the field strength
insertions are already higher order. A straight forward
calculation outlined in Appendix~\ref{ap:one-loop} leads to a term
in the action of the form
\begin{eqnarray}\label{eq:ntlo}
\int{\rm d}t{\rm d}^{2p}x_1{\rm d}^{2p}x_2\Big[c_1\rho_{FF}(x_1,t)\rho(x_2,t)+c_2\rho_F(x_1,t)\rho_F(x_2,t)\Big]\nonumber\\
\times\int\frac{{\rm d}\omega}{2\pi}
\widetilde{G}(\omega,\theta^{-1}x_{12})\widetilde{G}(\omega,\theta^{-1}x_{21}),\qquad\quad
\end{eqnarray}
where the subscript $F$ denotes an insertion of the field strength
into the end of the Wilson line.

At this point, there are several comments to be made. First of
all, it is clear that Eq.~(\ref{eq:ntlo}) describes the
instantaneous interaction between two points $x_1$ and $x_2$,
which is consistent with the long dipole picture depicted in
Fig.~\ref{fig:cyli}. Moreover, all one-loop matrix diagrams, which
differ only by higher dimensional operator insertions, must have a
similar double trace structure, and hence, give rise to two-body
interactions. Secondly, note that the appearance of the propagator
$\widetilde{G}$ is again consistent with our intuition from field
theory. In fact in Section~\ref{subsec:two}, we will see that this
property is a direct result of the gauge invariant structure of
perturbative corrections. Lastly, we can identify a term in
Eq.~(\ref{eq:ntlo}) that leads to one-loop renormalization of the
gauge coupling.

The one-loop renormalization comes from a UV divergence in the
planar sector,  corresponding to the constant part of $\rho$ in
the first term of Eq.~(\ref{eq:ntlo}). The integral over position
then factorizes into
\begin{eqnarray}\label{eq:coup}
 & & N\int{\rm d}t{\rm d}^{2p}x_1{\rho}_{FF}(x_1,t)\int\frac{{\rm d}\omega}{2\pi}{\rm d}^{2p}x_{12}\widetilde{G}({\omega},{\theta}^{-1}x_{12})^{2}\nonumber\\
 &=&  N\int{\rm d}t{\rm Tr}\left(\dot{B}^{i2}+[B^i,B^j]^2\right) \int\frac{{\rm d}\omega{\rm d}^{2p}p}{2\pi(2\pi)^{2p}}{\widetilde{G}(\omega,p)}^{2}.
\end{eqnarray}
This quantity is easily recognized as the familiar one-loop
contribution to the renormalization of the gauge coupling.
Although a systematic treatment of renormalization will have to
wait for Chapter~\ref{sec:renorm}, we can already see from
Eqs.~(\ref{eq:vacdiv}) and (\ref{eq:coup}) that UV dipoles in
planar diagrams can lead to UV divergences and conventional
renormalization of the parameters in the theory. Of course, we
shall assume a negative beta function so that our perturbative
Wilsonian approach is valid.

\section{Higher Loop Corrections to the Effective
Action}\label{subsec:two}

In the analysis of one-loop corrections to the quantum effective
action, we found that matrix diagrams naturally lead to
multi-trace operators, which had the physical interpretation of
instantaneous multi-body interactions mediated by long dipoles. We
will now extend our analysis to include higher order loop effects.
Ultimately, we will determine the general gauge invariant
structure of perturbation theory which will be exploited in the
Chapter~\ref{sec:renorm} to compute the divergence structure of
noncommutative gauge theory.

One of the entirely new features of higher order loops is the
appearance of nonplanar matrix diagrams.  For example, in the case
of the leading order two-loop diagrams, we have nonplanar cubic
and quartic diagrams
\begin{center}
\begin{minipage}[c]{1.75cm}
\includegraphics[scale=0.4]{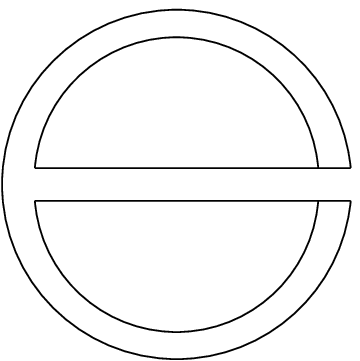}
\end{minipage}%
\begin{minipage}[c]{0.5cm}
$+$
\end{minipage}%
\begin{minipage}[c]{2.7cm}
\includegraphics[scale=0.4]{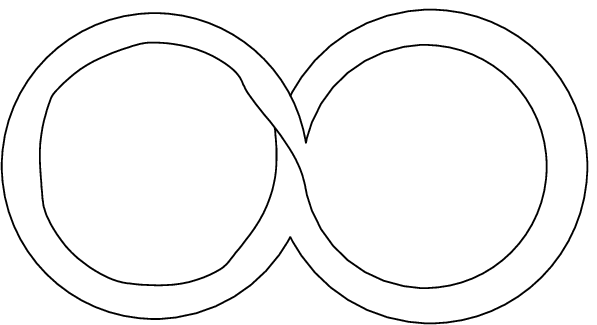}.
\end{minipage}
\end{center}
As discussed in  Appendix~\ref{ap:twoloop}, nonplanar matrix
diagrams correspond to field theory diagrams in which the loops
are linked in a nonplanar fashion.  However, nonplanar diagrams
seem to correspond to physics not described by the dipole degrees
of freedom that form the basis for our perturbative analysis.
Therefore, the validity of our analysis requires that we suppress
the contribution of nonplanar matrix diagrams to the Wilsonian
integration. It was shown in Appendix~\ref{ap:twoloop} that when
$\theta\Lambda^2\gg 1 $, nonplanar matrix diagrams are negligible
compared to planar ones due to exponential suppression from Moyal
phase factors. \emph{Thus, in the UV domain of Wilsonian
integration given by $\theta\Lambda^2\gg 1$, the contribution from
nonplanar matrix diagrams is negligible}.

\begin{figure}
\psfrag{p_0}{}\psfrag{x1}{$x_1$}\psfrag{x2}{$x_2$}\psfrag{x3}{$x_3$}
 \subfigure[An illustration of the contribution
from the first order treatment of quartic interaction terms given
by
(\ref{eq:threebody}).]{\label{quartic:subfig:a}\begin{minipage}[b]{.45\textwidth}\centering
\includegraphics[scale=0.5]{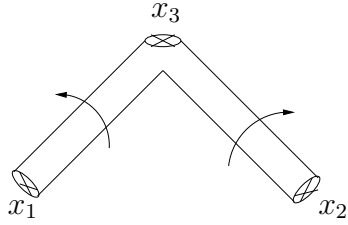}\end{minipage}}
\hspace{1cm} \subfigure[An illustration of the contribution  from
the second order treatment of cubic interaction terms given by
(\ref{eq:cubicthreebody}).]{\label{quartic:subfig:b}\begin{minipage}[b]{.45\textwidth}\centering
\includegraphics[scale=0.5]{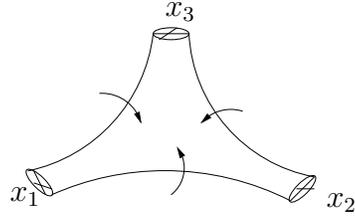}\end{minipage}}
\caption{Long distance three-body interactions corresponding to
high momentum dipoles propagating in two-loop
diagrams.}\label{fig:twolooppicture}
\end{figure}

Continuing with the leading order two-loop planar matrix diagrams
\begin{center}
\begin{minipage}[c]{1.75cm}
\includegraphics[scale=0.3]{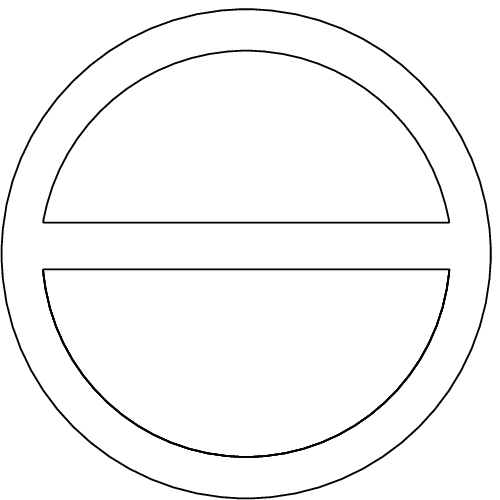}
\end{minipage}%
\begin{minipage}[c]{0.5cm}
$+$
\end{minipage}%
\begin{minipage}[c]{2.7cm}
\includegraphics[scale=0.4]{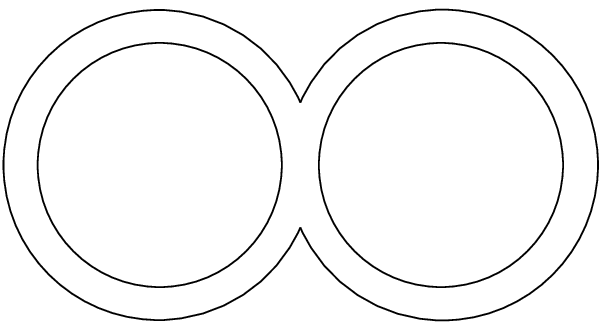},
\end{minipage}
\end{center}
we find triple trace contributions indicated by the double line
notation. As calculated in Appendix \ref{ap:twoloop}, the quartic
diagram gives an interaction term in the effective action
proportional to
\begin{eqnarray}\label{eq:threebody}
 & & \int{\rm d}t{\rm d}^{2p}x_1{\rm d}^{2p}x_2{\rm d}^{2p}x_3 \rho(x_1,t)\rho(x_2,t)\rho(x_3,t) \nonumber\\
 & & \qquad\times\int\frac{{\rm d}\omega_1}{2\pi}\frac{{\rm d}\omega_2}{2\pi}\widetilde{G}({\omega}_1,{\theta}^{-1}x_{13})\widetilde{G}({\omega}_2,{\theta}^{-1}x_{23}),
 \end{eqnarray}
and the cubic diagram gives a term  proportional to
\begin{eqnarray}\label{eq:cubicthreebody}
& & \int{\rm d}t{\rm d}^{2p}x_1{\rm d}^{2p}x_2{\rm
d}^{2p}x_3\rho(x_1,t)\rho(x_2,t)\rho(x_3,t)\nonumber\\
& & \qquad\times \int\frac{{\rm d}\omega_1}{2\pi}\frac{{\rm
d}\omega_2}{2\pi}(\omega_1\omega_2-x_{12}\cdot
x_{23})\widetilde{G}(\omega_1,\theta^{-1}x_{12})\nonumber\\
& & \qquad\qquad\qquad\qquad\times
\widetilde{G}(\omega_2,\theta^{-1}x_{23})
\widetilde{G}(\omega_1+\omega_2,\theta^{-1}x_{31}).
\end{eqnarray}
Indeed, as expected based on our intuition,
Eqs.~(\ref{eq:threebody}) and (\ref{eq:cubicthreebody}) describe
long-distance interactions that arise from high momentum dipoles
growing large in spatial extent and ``stretching out'' the matrix
diagrams, as depicted in Fig.~\ref{fig:twolooppicture}.
Furthermore, these expressions bear a close resemblance to what is
expected from ordinary field theory, and in fact, the general
structure of perturbative corrections is starting to emerge.

To each boundary of the double line diagram we should associate a
point in space and a trace which yields a gauge invariant Wilson
line. The position dependence of the interaction strength between
the Wilson lines follows from an integral over frequencies with
the integrand being given by a particular function of both the
dipole frequencies and momenta as defined by the separation
between the space points associated to the boundaries of the
double line diagram. In fact, the particular function of dipole
frequencies and momenta corresponds precisely to the structure of
momentum space field theory propagators that appear in the
analogous process in ordinary field theory.

The subleading terms in the derivative expansion must have a
similar trace structure but Wilson lines modified to include
higher derivative operator insertions as in Eq.~(\ref{eq:ntlo}).
Naturally, the subleading terms will include a different function
of dipole frequencies and momenta that reflects the extra
propagators that are required by the operator insertions. Since
powers of momentum are equivalent to powers of separation for
dipole degrees of freedom, the inclusion of more powers of
momentum in the denominator naively leads to faster falloff with
distance, which is expected from subleading terms in a derivative
expansion. However, quantum corrections coming from higher loop
orders are generally strong, as we will discuss in
Chapter~\ref{sec:renorm}. To further reinforce these ideas, we
have included an outline of the calculation of the next to leading
order two-loop planar matrix diagrams in Appendix~\ref{ap:ntlotl}.

Let us now exploit our understanding of the general gauge
invariant structure to obtain a result that holds to all orders of
perturbation theory. First of all, it is clear that at $L$-loop
order the leading interactions will involve $L+1$ Wilson lines
with no operator insertions, such as Eq.~(\ref{eq:trac}) in the
case $L=1$ and both Eq.~(\ref{eq:threebody}) and
Eq.~(\ref{eq:cubicthreebody}) in the case $L=2$. The crucial
observation is that since the vacuum diagrams are contained in
interactions of this form, as shown explicitly by
Eq.~(\ref{eq:vacdiv}) in the one-loop case, the leading order
terms must vanish entirely if the vacuum diagrams vanish. For
example, this result implies that for supersymmetric
noncommutative theories the leading interactions must cancel at
each order in perturbation theory. This statement is a
generalization to all loop orders of the cancellation that we have
already seen in the case of the leading one-loop interactions
given by Eq.~(\ref{eq:trac}). It reflects the fact that
supersymmetric theories are softer in the UV, and hence, the IR
behavior that is generated by UV-IR mixing is not as strong. The
relation between the amount of supersymmetry and the cancellation
of nonanalytic terms in external momenta has been discussed in
\cite{vv:wilson,ls:gaug} from the field theory perspective.

Actually, we can extend this line of reasoning much farther. In
order to do so, let us take some time to discuss the
proportionality constants that we have thus far ignored. Consider
the content of the matrix diagrams from a field theory point of
view. First of all, the double line matrix diagrams themselves
only contain information pertaining to the contraction of gauge
indices, which encodes the planarity and nonplanarity of both the
external operator insertions and the internal propagators. Thus,
the double line diagrams contain purely topological information,
not specific to any particular gauge theory -- ordinary or
noncommutative. In terms of the framework that we have discussed
up to this point, this topological information only fixes the
ratio of coefficients between terms with a similar propagator
structure but different Wilson line structure. For example, in
Eq.~(\ref{eq:ntlo}) the ratio between $c_1$ and $c_2$ is fixed by
purely topological considerations. It also follows that the
coefficients of nonplanar matrix diagrams are fixed relative to
the coefficients of the corresponding planar diagrams (Note that
factors of $N$ result from the traces that are included in the
Wilson lines and loops and are not contained in the coefficients
as we have defined them). The essential point is that topological
quantities, such as the ratio between coefficients of diagrams
involving a similar propagator structure, are independent of most
characteristics specific to the theory in question such as the
field content, coupling constants, and whether the theory is
noncommutative or not.  Topological quantities can only depend on
properties that determine how  double lines can be ``glued"
together to form diagrams.

Nonetheless, the coefficients of matrix diagrams involving a
similar propagator structure all share a common normalization
factor that depends on the number and type of modes that propagate
around the loops. Thus, the overall normalization factor does
depend on characteristics specific to a particular theory, such as
the field content and the coupling constants. However, the
noncommutativity of the theory enters only through the Moyal phase
factors inside the momentum integrals of nonplanar field theory
diagrams; noncommutativity does not affect the coefficient
multiplying the integral. After all, the combinatorics associated
with the contraction of fields is the same in both cases.
\emph{Thus, the coefficients of double line matrix diagrams in a
given noncommutative gauge theory are equal to the corresponding
coefficients of double line diagrams in the ordinary counterpart
gauge theory}.

It follows immediately that the nonrenormalization theorems
enjoyed by ordinary supersymmetric theories generalize to
noncommutative supersymmetric theories.  For example, we already
know that even minimal supersymmetry leads to the vanishing of the
leading order matrix diagrams -- both planar and nonplanar -- to
all orders in perturbation theory. However, we can now make some
much stronger statements. For example, by imposing $3+1$
dimensional $\mathcal{N}=2$ supersymmetry, we can eliminate all
corrections to the gauge coupling beyond one loop. In this case,
the next to leading order one-loop correction given by
Eq.~(\ref{eq:ntlo}) is not constrained, but supersymmetry does
impose restrictions at the two-loop level. As discussed in
Appendix~\ref{ap:ntlotl}, $b_6$ contains the two-loop correction
to the beta function, which must vanish. Since $b_7$ and $b_8$ are
also proportional to $b_6$ by the arguments above,
Eq.~(\ref{eq:dangerous}) must vanish entirely for $3+1$
dimensional $\mathcal{N}=2$ noncommutative Super Yang-Mills
(NCSYM) theory. In the case of $3+1$ dimensional $\mathcal{N}=4$
NCSYM theory, none of the next to leading order interactions are
allowed at all.

In summary, we have shown how dipole degrees of freedom naturally
emerge from a manifestly gauge invariant perturbative treatment of
noncommutative gauge theories in the UV regime defined by
$\theta\Lambda^2\gg 1$. The essential result of this Chapter,
however, was the determination of the gauge invariant form of
perturbative corrections to the Wilsonian quantum effective
action. In particular, we were able to relate the proportionality
constants of double line matrix diagrams that appear in
noncommutative gauge theories to the corresponding constants in
ordinary gauge theories, with which we are familiar and many
results are known. In Chapter~\ref{sec:renorm}, we shall exploit
this isomorphism to its fullest extent in order to make powerful
statements about the divergence structure of noncommutative gauge
theories.

\chapter{Application of the Perturbative Wilsonian Formalism to
the Divergence Structure}\label{sec:renorm}
\index{Making the Bibliography with BiBTeX%
@\emph{Making the Bibliography with BiB\TeX{}}}%

\section{Introduction}

We will now systematically analyze the divergence structure of the
perturbative corrections to the Wilsonian quantum effective
action. In the process of integrating out UV modes, in addition to
UV divergences, we encounter new divergences having dual UV and IR
interpretations. Using the structural results discussed in
Chapter~\ref{sec:pert}, we are able to show that, to all orders in
perturbation theory, the pure UV divergences can be cancelled by
standard renormalization of the parameters in the theory, but the
UV-IR divergences can only be cancelled by adding terms of an
entirely new form to the action. We then discuss self-consistency
conditions on noncommutative gauge theories and nonrenormalization
theorems that prevent the appearance of UV-IR divergences. We also
briefly discuss IR divergences that arise from the naive
perturbative treatment of the quantum effective action.

\section{Divergences and UV-IR Mixing}\label{subsec:survey}

In Section~\ref{subsec:one}, we calculated the one-loop
divergences. They were shown to be of the standard UV form and
merely contribute to the renormalization of the vacuum energy and
the gauge coupling. However, we shall now show that the effect of
UV-IR mixing on the divergence structure enters at the two-loop
level and qualitatively changes the types of divergences that can
appear. There are pure UV divergences, dual UV-IR divergences, and
pure IR divergences.

Let us start by discussing the IR divergences.  These divergences
do not arise directly from the Wilsonian integration of the UV
modes, but rather, arise from the integration of IR modes in the
context of a naive perturbative treatment of the quantum effective
action. The most primitive two-loop example comes from the
interactions given by Eq.~(\ref{eq:trac}). To see this, it is
convenient to Fourier transform to momentum space, in which case
we get
\begin{equation}
\int{\rm d}t\frac{{\rm d}\omega{\rm
d}^{2p}p}{(2\pi)(2\pi)^{2p}}\frac{ {\rm
d}^{2p}k}{(2\pi)^{2p}}e^{ip\cdot\theta\cdot
k}\tilde{\rho}(k,t)\tilde{\rho}(-k,t)\log\widetilde{G}(\omega,p).
\end{equation}
We now perturbatively expand the Wilson lines
\begin{equation}
\tilde{\rho}(k,t)=N(2\pi)^{2p}\delta^{2p}(k)+ik\cdot\theta\cdot
{\rm tr}_N\tilde{A}(k,t)+\ldots,
\end{equation}
and contract the terms linear in the gauge field.  The result is
proportional to
\begin{eqnarray}\label{eq:irdiv1}
& & NV\int\frac{{\rm d}\omega {\rm
d}^{2p}p}{(2\pi)(2\pi)^{2p}}\frac{{\rm d}\omega^\prime {\rm
d}^{2p}k}{(2\pi)(2\pi)^{2p}}k^2e^{ip\cdot\theta\cdot
k}\widetilde{G}(\omega^\prime,k)\log\widetilde{G}(\omega,p),
\end{eqnarray}
where the UV loop momentum $p>\Lambda$ and the IR loop momentum
$k<\Lambda$. To recast Eq.~(\ref{eq:irdiv1}) into a more familiar
form, we must rewrite
\begin{equation}
k^2e^{ip\cdot\theta\cdot
k}=-ik\cdot\theta^{-1}\cdot\partial_p\left(e^{ip\cdot\theta\cdot
k}\right),
\end{equation}
and integrate by parts to obtain
\begin{eqnarray}\label{eq:irdiv2}
NV\int\frac{{\rm d}\omega {\rm
d}^{2p}p}{(2\pi)(2\pi)^{2p}}\frac{{\rm d}\omega^\prime {\rm
d}^{2p}k}{(2\pi)(2\pi)^{2p}}ik\cdot\theta\cdot
pe^{ip\cdot\theta\cdot
k}\widetilde{G}(\omega^\prime,k)\widetilde{G}(\omega,p).
\end{eqnarray}
Again, we manipulate Eq.~(\ref{eq:irdiv2}) by rewriting
\begin{equation}
-ik\cdot\theta\cdot pe^{ip\cdot\theta\cdot k}=\partial_\sigma
e^{i\sigma p\cdot\theta\cdot k}\big|_{\sigma=1}.
\end{equation}
After re-scaling the frequencies and momenta, it is easy to see
that Eq.~(\ref{eq:irdiv2}) reduces to the familiar form of the IR
divergent two-loop nonplanar vacuum diagram whose contribution is
given by Eq.~(\ref{eq:exp}). Therefore, this well-known IR
divergence that is ubiquitous in noncommutative quantum field
theories is actually contained in the contribution to the
effective action of the one-loop planar matrix diagram. In fact,
it is clear that in our Wilsonian framework, IR divergent
nonplanar diagrams always appear from the contraction of external
insertions on distinct boundaries of UV divergent planar matrix
diagrams. The reason is that, due to Moyal phase factors,
nonplanar loop diagrams contribute only when some loop momenta are
in the UV, while others are in the IR. Thus, the existence of IR
divergences in nonplanar loop diagrams is directly related to the
existence of UV divergences in planar matrix diagrams.

It follows that the IR divergences can be cancelled by any
mechanism that cancels the UV divergent planar diagrams, such as
minimal supersymmetry in $2+1$ dimensions or maximal supersymmetry
in $3+1$ dimensions. Although, it should be noted that it is
possible that the IR divergences do not need to be cancelled.
Perhaps IR divergences can be dealt with by reorganizing the
perturbative expansion in some manner. For example, the
resummation of the leading nonanalytic momentum dependence into
the propagator has been proposed in \cite{ms:nonp}, but this
technique only works for theories containing a matter content such
that the leading nonanalytic term has the proper sign for the
summation not to lead to an instability \cite{ar:uvir,mv:mean}.
Other authors have studied the IR behavior of noncommutative gauge
theories in the context of the star product formulation of
Wilsonian integration \cite{vv:wilson}. However, it could well be
that a deeper structural understanding of the theory is required
in order to resolve the problem of IR divergences. In fact, it may
turn out that the weakly coupled dipoles that are valid for
$\theta\Lambda^2\gg 1$ may not be the correct degrees of freedom
when $\theta\Lambda^2\lesssim 1$ and nonplanar matrix diagrams
become important. In any event, we expect that the gauge invariant
Wilson loop structure associated with nonplanar diagrams will play
a prominent role in resolving this issue.

A further discussion of the systematic treatment of IR divergences
is beyond the scope of this work. Instead, we seek to understand
the divergences that emerge directly from the Wilsonian
integration of UV modes. Let us consider the leading order
two-loop diagrams, which can be combined into the form of
Eq.~(\ref{eq:threebody}). It is now straight forward to isolate
the divergent terms by using the splitting scheme $\rho(x)={\rm
tr}_{N}({\rlap{1} \hskip 1.6pt \hbox{1}})+\Delta(x)$ that was
discussed in Section~\ref{subsec:one}. The constant term in either
$\rho(x_1)$ or $\rho(x_2)$ gives
\begin{equation}\label{eq:propcorrection}
 \int{\rm d}t{\rm d}^{2p}x_1{\rm d}^{2p}x_3 \rho(x_1,t)\rho(x_3,t)\int \frac{{\rm d}\omega_1}{2\pi}\widetilde{G}(\omega_1,{\theta}^{-1}x_{13})\int\frac{{\rm d}\omega_2{\rm
d}^{2p}p_2}{2\pi(2\pi)^{2p}}\widetilde{G}(\omega_2,p_2).
\end{equation}
Apparently, Eq.~(\ref{eq:propcorrection}) involves the one-loop UV
divergent contribution to the mass term of the field theory
propagator $\widetilde{G}$ appearing in Eq.~(\ref{eq:trac}), which
is a familiar type of correction from our experience with ordinary
quantum field theories. Of course, gauge invariance prevents these
type of corrections to the gauge field propagator, and more
generally, supersymmetry can be employed to cancel them in all
cases. Nonetheless, the meaning of Eq.~(\ref{eq:propcorrection})
is that it represents a one-loop correction to the one-loop
interaction given by Eq.~(\ref{eq:trac}); it does not contain an
inherently two-loop correction to the theory. Furthermore, the
diagrammatical interpretation of Eq.~(\ref{eq:propcorrection}) is
that it contains the contribution from a one-loop planar
subdiagram, which necessarily involves a loop integration that is
decoupled from the Moyal phase factors. Since Moyal phase factors
are the source of UV-IR mixing, this means that divergences in
planar subdiagrams always have a purely UV interpretation.

The other distinct group of terms in the contribution of the
leading order two-loop diagrams comes from the constant term of
$\rho(x_3)$
\begin{equation}\label{eq:twoloopcorrection}
  \int{\rm d}t{\rm d}^{2p}x_1{\rm d}^{2p}x_2 \rho(x_1,t)\rho(x_2,t)\int\frac{{\rm d}\omega_1}{2\pi}\frac{{\rm d}\omega_2 {\rm
d}^{2p}p_3}{2\pi(2\pi)^{2p}}\widetilde{G}(\omega_{21},p_3-{\theta}^{-1}x_{12})\widetilde{G}(\omega_2,p_3).
\end{equation}
First, we consider the UV divergent contributions from the planar
subdiagrams contained in Eq.~(\ref{eq:twoloopcorrection}). The
leading order contribution from planar non-vacuum diagrams
vanishes, as usual, because $\int {\rm d}^{2p}x\Delta(x)=0$.
However, the vacuum diagrams, which are contained in the constant
terms of $\rho(x_1)$ and $\rho(x_2)$, give the two-loop correction
to the renormalization of the vacuum energy. This is to be
contrasted with the vacuum diagrams contained in
Eq.~(\ref{eq:propcorrection}), which give the mass correction to
the propagators in the one-loop vacuum diagrams. In fact, this
example shows how the structure added by the background insertions
removes all ambiguities associated with overlapping divergences.

The nonplanar diagrams contained in
Eq.~(\ref{eq:twoloopcorrection}), corresponding to  the $\Delta^2$
term, evidently represent a two-loop quantum correction to the
interaction strength of Eq.~(\ref{eq:non1}). The growth in
$|x_1-x_2|$ of this correction is given by
\begin{eqnarray}\label{eq:intcorrection}
\int\frac{{\rm d}\omega_1}{2\pi}\frac{{\rm d}\omega_2 {\rm
d}^{2p}p_3}{2\pi(2\pi)^{2p}}\widetilde{G}(\omega_{21},p_3-{\theta}^{-1}x_{12})\widetilde{G}(\omega_2,p_3),
\end{eqnarray}
which is clearly divergent for $p\geq 1$. In order to understand
the meaning of divergent loop diagrams with nonplanar external
insertions, such as those contained in
Eq.~(\ref{eq:twoloopcorrection}), it is instructive to look at the
interaction in momentum space
\begin{equation}\label{eq:moyal}
 \int{\rm d}t\frac{{\rm d}^{2p}k}{(2\pi)^{2p}}\widetilde{\Delta}(k,t)\widetilde{\Delta}(-k,t) \frac{{\rm d}\omega_1{\rm d}^{2p}q}{2\pi(2\pi)^{2p}} \frac{{\rm d}\omega_2{\rm d}^{2p}p}{2\pi(2\pi)^{2p}}e^{ik\cdot\theta\cdot
q}\widetilde{G}(\omega_{21},p-q)\widetilde{G}(\omega_2,p),
\end{equation}
because the role of the Moyal phase factors is clarified. We can
first perform the integrals over $\omega_2$ and $p$ by introducing
a Schwinger parameter and an UV cutoff $M$, as in \cite{ms:nonp}.
Then the integral over $\omega_1$ and $q$ is regular, and we are
left with a quantity proportional to
\begin{eqnarray}
& & \int{\rm d}t\frac{{\rm
d}^{2p}k}{(2\pi)^{2p}}\frac{\widetilde{\Delta}(k,t)\widetilde{\Delta}(-k,t)}{(
k^{2}+1/{M^{2}})^{2p-1}}\nonumber\\
&=& \int{\rm d}t{\rm d}^{2p}x_1{\rm d}^{2p}x_2
\Delta(x_1,t)\Delta(x_2,t)\int\frac{{\rm
d}^{2p}k}{(2\pi)^{2p}}\frac{e^{ik\cdot(x_1-x_2)}}{(
k^{2}+1/{M^{2}})^{2p-1}}.
\end{eqnarray}
Now the $M\rightarrow\infty$ limit results in a divergence in the
IR region of integration in a Fourier transform instead of the UV
region of integration in a loop integral, and $1/M$ plays the role
of an IR cutoff. The transform back into position space yields a
leading order interaction strength of the form
$|x_1-x_2|^{2p-2}\log(|x_1-x_2|^2/{M^2})$, which could also be
determined directly from Eq.~(\ref{eq:intcorrection}).

Apparently, divergences in loop diagrams with nonplanar external
insertions have a dual UV-IR interpretation. In the case of
Eq.~(\ref{eq:moyal}), the origin of this duality can be traced to
the Moyal phase factor $\exp(ik\cdot\theta\cdot q)$, which
represents the nonplanarity of the background insertions and gives
rise to UV-IR mixing. In particular, the UV loop integrals over
virtual states, labelled by momenta $p$ and $q$, produce
nonanalytic dependence on the background momentum $k$ that becomes
important in the IR. In other words, the divergence comes from
integrating over UV states, but on the other hand, it can be
recast into the form of an IR singular Fourier transform.

Actually, this type of UV-IR mixing has a simple interpretation in
terms of dipole degrees of freedom. The separation between the
endpoints of the dipoles is the dual variable to the external
momenta $k$. Therefore, the IR singular Fourier transform gives
rise to corrections to interaction strengths which grow strong
with large separation. However, since powers of separation are
equivalent to powers of dipole momentum, these strong corrections
can also be thought of as arising from UV divergences in the
theory. Thus, the dipole intuition that emerges from the matrix
formulation seems to shed new light on the proposal of
\cite{cm:dual} concerning the dual interpretation of divergences
in nonplanar diagrams, although our treatment of UV-IR divergences
will be much different than theirs.

Although we have surveyed only the contribution from the leading
order two-loop diagrams explicitly, it is straight forward to
perform a similar analysis on the next to leading order two-loop
diagrams that were calculated in Appendix~\ref{ap:ntlotl}. It is
easy to see that the next to leading order divergence structure is
qualitatively similar. In fact, the divergence structure of all
the subleading terms in the derivative expansion will be
qualitatively similar to the examples that we have discussed
explicitly, although in $3+1$ dimensions or fewer, power counting
implies that there are no divergences of any kind beyond the next
to leading order. However, it remains to be seen what the effect
of higher order loop corrections will mean for the
quantum-mechanical self-consistency of noncommutative gauge
theories.

\section{Treatment of UV and UV-IR
Divergences}\label{subsec:higher}

In Section~\ref{subsec:survey}, we gained a two-loop introduction
to the types of divergences that appear in noncommutative
theories. We found pure UV divergences, dual UV-IR divergences,
and pure IR divergences. While the proper treatment of the IR
divergences lies beyond the scope of this work, we will be able to
understand the structure of both the UV and UV-IR divergences.
Ultimately this divergence structure follows from the gauge
 invariant form of perturbative corrections that we determined in
Section~\ref{subsec:two} and the diagrammatical combinatorics,
which we shall discuss below.

As we have seen explicitly at the one- and two-loop levels, the
divergent loop integrations appearing in planar subdiagrams have a
pure UV interpretation and are structurally consistent with the
renormalization of parameters such as the vacuum energy, mass, and
gauge coupling. In fact, it is easy to see that the UV divergence
structure is consistent with parameter renormalization to all loop
orders. Moreover, there are no ambiguities posed by overlapping
divergences due to the structure added by the external background
insertions, which are automatically included in this formalism.
However, it is not yet clear whether the combinatorics of the
planar subdiagrams is consistent with renormalizability to all
orders in perturbation theory.

Actually, it turns out that the diagrammatical combinatorics of
noncommutative theories is isomorphic to that of the ordinary
counterpart theories. To see this, it is helpful to consider the
contribution from matrix diagrams -- at any order in the
derivative expansion, planar, or nonplanar -- in energy-momentum
space. For example, Eq.~(\ref{eq:trac}) becomes
\begin{equation}\label{eq:combo}
\int\frac{{\rm d}^{2p+1}p}{(2\pi)^{2p+1}}\frac{{\rm
d}^{2p+1}k}{(2\pi)^{2p+1}}e^{ip\cdot\theta\cdot
k}\tilde{\rho}(k)\tilde{\rho}(-k)\log\widetilde{G}(p),
\end{equation}
where
$\tilde{\rho}(k)=N(2\pi)^{2p+1}\delta^{2p+1}(k)+\widetilde{\Delta}(k)$.
Clearly, the contribution from any matrix diagram can be written
this way by Fourier transforming the Wilson lines and loops and
interpreting the separation between the endpoints of the virtual
dipoles as a momentum variable instead of a separation variable.
The point is that, other than the Moyal phase factor and the
particular form of $\widetilde{\Delta}(k)$, Eq.~(\ref{eq:combo})
it identical to what is expected from ordinary gauge theory.
Furthermore, this statement is true for any matrix diagram
whatsoever -- the interpretation is that $\widetilde{\Delta}(k)$
encodes the gauge invariant contribution of background insertions
of net momentum $k$ into the boundary of the diagram; while the
delta function term reduces to the contribution from diagrams with
no insertions on the corresponding boundary.

The crucial observation is that the diagrammatical combinatorics,
which follows from expanding $\tilde{\rho}$ as above, is
completely independent of the Moyal phase factors and the
particular form of $\widetilde{\Delta}$. Therefore, the
diagrammatical combinatorics of a particular noncommutative gauge
theory is absolutely identical to that of the ordinary
counterpart, which we shall assume to be renormalizable. Since we
know that renormalizability can be understood in the context of
the $1/N$ expansion, which separates planar and nonplanar
contributions, it follows immediately that the UV divergences that
occur in planar subdiagrams of noncommutative theories can be
cancelled by parameter renormalization to all orders in
perturbation theory. It does not matter that nonplanar diagrams
are no longer UV divergent when the theory is noncommutative; it
only matters that the combinatorics are such that the divergences
in the planar subdiagrams can be cancelled by parameter
renormalization.  \emph{Thus, renormalizability of UV divergences
is a property inherited by noncommutative gauge theories from
their ordinary counterparts}.

It follows that a necessary condition for the removal of the UV
cutoff in a given noncommutative gauge theory is that the ordinary
counterpart gauge theory be renormalizable. Therefore, we can only
consider (spatially) noncommutative theories living in either
$2+1$ or $3+1$ dimensions, if we demand that the theory be
self-consistent quantum-mechanically. Of course, this was to be
expected based on our intuition from ordinary quantum field
theories. However generally, there are still UV-IR divergent terms
which appear as corrections to the strength of lower order
long-distance interactions, which are described by purely
nonplanar diagrams. In fact, we can reduce the divergence
structure down to purely nonplanar form by extracting all of the
UV divergences from planar subdiagrams via renormalization. Then,
any remaining divergence must come from purely nonplanar diagrams,
and therefore, must be of the UV-IR type.

Actually, as shown explicitly at the two-loop order in
Section~\ref{subsec:survey}, the existence of UV-IR divergences is
directly related to the existence of higher loop order corrections
to the renormalization of the theory. The reason is that the UV-IR
divergences that are new at a given loop order can only come from
the group of terms containing the UV divergent planar diagrams
that contribute a higher loop order correction to renormalization,
because all other planar subdiagrams simply amount to lower order
loop corrections. Put another way, UV-IR divergent nonplanar
diagrams correspond precisely to the contribution from otherwise
planar subdiagrams  that would give higher order corrections to
renormalization were it not for extra nonplanar external
insertions. Therefore, UV-IR divergences will generally appear in
noncommutative theories that allow renormalization beyond the
one-loop order.

Actually, there is an accidental exception to this rule which is
due to $g^2$ having mass dimension one in $2+1$ dimensions. To see
this, we need only consider the leading order terms in the
derivative expansion, since only they allow divergences in this
case. Furthermore, it is easy to see that the two-body $\Delta^2$
interactions
\begin{equation}\label{eq:loopcorr}
\int{\rm d}t{\rm d}^2 x_1{\rm d}^2 x_2
\Delta(x_1,t)\Delta(x_2,t)\Big(|x_1-x_2|+\textrm{higher loop
corrections}\Big),
\end{equation}
are the only dangerous terms. Based on dimensional analysis, the
two-loop term in Eq.~(\ref{eq:loopcorr}) is of the form
$g^2\log(|x_1-x_2|^2/M^2)$, which is also consistent with
Eq.~(\ref{eq:intcorrection}). However, in this case, the actual
logarithmic divergence does not contribute due to $\int{\rm
d}x\Delta(x)=0$. Moreover, three-loop and higher corrections are
finite since the degree of divergence is lower at each order.
Therefore, in $2+1$ dimensions, UV-IR divergences do not appear.
Of course, this same argument does not work in $3+1$ dimensions
because there are new logarithmic divergences at each order in
perturbation theory. In the particular case of
Eq.~(\ref{eq:loopcorr}), the higher loop corrections in $3+1$
dimensions are of the form of powers of the dimensionless quantity
$g^2\log(|x_1-x_2|^2/M^2)$ multiplying the one-loop term, which
goes as $|x_1-x_2|^2\log(|x_1-x_2|^2/M^2)$.

\emph{Thus, UV-IR divergences occur in noncommutative gauge
theories of dimension $3+1$ or higher if and only if
renormalization beyond one loop is allowed}. Furthermore, these
UV-IR divergences cannot be cancelled by parameter
renormalization, because structurally they contribute entirely new
terms to the action. For this reason, quantum-mechanical
self-consistency requires that the appearance of UV-IR
divergences, and hence renormalization beyond one loop, is not
allowed. Of course, nonrenormalization properties of this type
require supersymmetry, which is simply a statement that softer UV
behavior is necessary to control the proliferation of UV-IR
divergences arising from UV-IR mixing. In fact, it is interesting
that in $3+1$ dimensions minimal $\mathcal{N}=1$ supersymmetry is
not sufficient; at least $\mathcal{N}=2$ is required to protect
against renormalization beyond one-loop order, and hence, UV-IR
divergences.

One must keep in mind, however, that the consistent removal of the
UV cutoff does not necessarily imply the theory is well behaved in
the IR. In fact, our analysis in Section~\ref{subsec:survey}
implies that if UV divergences are allowed at all, IR divergences
will appear in the naive perturbative treatment of the Wilsonian
quantum effective action. For example, $3+1$ dimensional
$\mathcal{N}=2$ NCSYM, which allows only a one-loop
renormalization of the gauge coupling, contains IR divergences in
the next to leading order nonplanar diagrams such as
\begin{center}
\includegraphics[scale=.4]{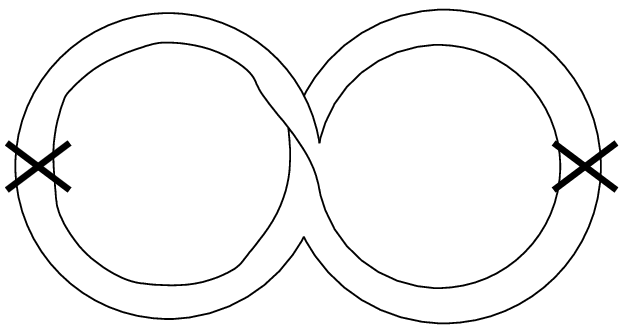}.
\end{center}
In $2+1$ dimensions, IR divergences generally appear in the
leading order two-loop nonplanar diagram. In fact, the only way to
eliminate these IR problems completely in the context of the
conventional perturbative expansion discussed in this work, is to
demand at least minimal supersymmetry in $2+1$ dimensions and
maximal supersymmetry in $3+1$ dimensions. As discussed in
Section~\ref{subsec:survey}, it remains an interesting and
important open problem as to whether or not this is necessary for
consistency.

On another note, it is tempting to extend our results to
noncommutative theories other than gauge theories. After all, the
intuition that we have gained appears to be quite generic and most
likely applies to any noncommutative theory. For example, it seems
unlikely that $3+1$ dimensional noncommutative scalar theory is
self-consistent quantum-mechanically, when nonsupersymmetric $3+1$
dimensional gauge theories are not. It even seems reasonable that
self-consistency could break down in $\mathcal{N}=1$ and
$\mathcal{N}=2$ supersymmetric noncommutative theories that do not
include gauge degrees of freedom, in light of the fact that any
extra hypermultiplets of matter that are added to $\mathcal{N}=2$
NCSYM theory must form particular representations which do not
allow all orders of wave function renormalization.

Nonetheless, arguments have been put forward for the quantum
consistency of many noncommutative theories that are excluded by
our analysis \cite{cr:renorm}. Although, these works have involved
different approaches which have encoded UV-IR mixing in one way or
another, none have employed a manifestly dipole construction such
as the matrix formulation. Since the dipole behavior of the
elementary quanta is the fundamental origin of UV-IR mixing in
noncommutative theories, it seems that the physical interpretation
that naturally emerges from the matrix approach is the most
reliable. We believe that this is a tremendous advantage when
discussing the divergence structure, because traditionally, the
proper treatment of divergences has resulted from a sound physical
interpretation for their meaning. Of course, the physical content
of noncommutative theories is independent of the language used to
discuss them; we are simply suggesting that the physics is more
clear in the matrix representation.

Ultimately, the essential difference between our work and
\cite{cr:renorm} is the interpretation and treatment of the dual
UV-IR divergences that occur in noncommutative quantum theories.
Other authors have employed independent UV and IR cutoffs in their
analysis. They have found that the UV divergences can be
renormalized and the UV cutoff removed while the UV-IR and IR
divergences  are cutoff independently in the IR (Actually, they do
not distinguish between UV-IR and IR divergences). Of course, this
result is not inconsistent with our analysis, but at some point
the IR cutoff must be removed in which case the UV-IR divergences
reappear. In other words, by independently cutting off the IR,
they inadvertently hide the UV-IR divergences which actually come
directly from the UV region of integration, and therefore, must be
dealt with before the UV cutoff is removed. In fact, independent
UV and IR cutoffs are not even consistent with UV-IR mixing
because UV and IR degrees of freedom cannot be separately removed
from the theory due to their mutual nondecoupling. Thus, the
results of \cite{cr:renorm} may not be entirely reliable.

Before closing, let us make a few comments concerning the scope of
our results from the perspective of string theory. Our analysis
has been limited to noncommutative gauge theories, and given these
dipole degrees of freedom, we have shown that both a sufficiently
high degree of supersymmetry and low spacetime dimensionality is
necessary to ensure the quantum mechanical consistency of the
theory. However, this result does not imply that other degrees of
freedom cannot be added to the theory to give a consistent
completion. For example, \cite{rg:ncos} shows that there is an UV
completion of some higher dimensional NCSYM theories in the form
of noncommutative open string theories, in which the closed string
sector has decoupled but there are still stringy modes from the
open string sector that remain. Yet another distinct possibility
is that, in the case of theories with a lesser degree of
supersymmetry, there could be some closed string modes that
survive the decoupling limit and render the theory consistent
\cite{ar:closed}. The point is that any theory that emerges from
an exact decoupling limit of string theory must be consistent; our
results only imply that the decoupled theory can conceivably be a
noncommutative gauge theory in only very special cases.

\chapter{Discussion and Outlook}\label{sec:conc}
\index{Making Tables and Including Figures@\emph{Making Tables
    and Including Figures}}%

In this work, we have developed the perturbative Wilsonian
treatment of noncommutative gauge theories in the matrix
formulation. These methods proved to be instrumental in
understanding the effects of UV-IR mixing. In particular, we were
able to make significant progress in two main respects.

First, we were able to directly observe the fundamental dipole
structure of the elementary quanta. These dipoles had the property
that they extended proportionately in the transverse direction to
their center of mass momentum, which is expected from a string
theory analysis of the decoupling limit that leads to
noncommutative gauge theory. Furthermore, we found that the dipole
character emerged naturally from the matrix representation and was
manifestly embodied by the propagator.  When UV states were
integrated out, the perturbative corrections to the Wilsonian
quantum effective action included nonlocal interaction terms that
were structurally consistent with the dipole behavior of the
intermediate virtual quanta. In the end, UV-IR mixing could be
understood as originating from the dipole nature of the theory: UV
dipoles grow long in spatial extent and mediate instantaneous
long-distance interactions that are relevant in the IR.

Secondly, we found that UV-IR mixing has a profound effect on the
divergence structure of the theory. In the process of integrating
out UV modes, we encountered divergences having pure UV and dual
UV-IR interpretations. The pure UV divergences were shown to
contribute to the renormalization of parameters in the theory when
the dimensionality does not exceed $3+1$. On the other hand, the
dual UV-IR divergences, which were shown to appear in
dimensionality greater than $2+1$, could not be cancelled by
parameter renormalization. Instead, we had to invoke a minimum of
$\mathcal{N}=2$ supersymmetry in $3+1$ dimensions. Furthermore, it
was shown that a naive perturbative analysis of the Wilsonian
quantum effective action results in IR divergences in any
dimensionality. The general treatment of these IR divergences
required a structural understanding beyond the scope of this work;
although, it was shown that minimal supersymmetry in $2+1$
dimensions and maximal supersymmetry in $3+1$ dimensions were
sufficient to eliminate all IR divergences from the theory.

The main tool that we employed in our analysis was noncommutative
gauge invariance. The understanding of the gauge invariant
structure of perturbative corrections to the quantum effective
action enabled us to extend our analysis to all orders of
perturbation theory, both in terms of what nonlocal interactions
and divergences are allowed to appear.  Moreover, the gauge
invariant structure, particularly in regards to the connection
between Wilson loops and nonplanar diagrams, is likely to play a
central role in understanding the IR regime of Wilsonian
integration in noncommutative gauge theories. In particular, it is
not even known how to describe the IR degrees of freedom that are
valid when $\theta\Lambda^2\lesssim 1$; our analysis, based on
dipole degrees of freedom, is only reliable in the UV regime given
by $\theta\Lambda^2\gg 1$. Nonetheless, we expect that the IR
degrees of freedom receive large contributions from the UV, and
are likely described by manifestly nonlocal objects such as Wilson
lines and loops. Understanding the IR regime of Wilsonian
integration is perhaps the most interesting and important problem
that remains unsolved in the subject of noncommutative gauge
theories.

However, as discussed in the introduction, there are more
fundamental reasons to understand the role of noncommutative gauge
invariance. While the preliminary steps in analyzing the behavior
of noncommutative gauge theories have not obviously shed much
light on either duality or holography, we have been able to gain
some intuition for how nonlocal gauge invariance (not to be
confused with invariance under global symmetries) can give rise to
spatially extended degrees of freedom that exhibit UV-IR mixing.
Both of these ingredients are necessary in any holographic theory,
so it is possible that some hint of progress has been made in this
direction. Duality, on the other hand, does not seem to be
illuminated at all by our  study of noncommutative gauge symmetry,
although it is possible that some sort of duality emerges between
the degrees of freedom in the UV and those in the IR
\cite{sj:dual}.

Nevertheless, this work is significant in the sense that it
contributes to the development of noncommutative gauge theories
themselves.  Fundamental physics aside, noncommutative gauge
theories are interesting in their own right, mainly because they
exhibit manifest spatial nonlocality and UV-IR nondecoupling. Both
of these properties have posed major challenges to the
understanding of noncommutative quantum field theories. This work
has shed a great deal of light on how to formulate and interpret
the physics of such theories, as well as open the door for future
studies in the area.  We look forward to finding out where this
line of research will ultimately lead.

%
%
\appendices
\index{Appendices@\emph{Appendices}}%

\chapter{Orthogonality and Completeness Relations for the Fourier Matrix
Basis}\label{ap:orthocomp}
\index{Appendix!Lerma's Appendix@\emph{Lerma's Appendix}}%

In this Appendix, we will derive the orthogonality and
completeness relations for the Fourier matrix basis discussed in
Section~\ref{subsec:matrix}. Our approach will be to explicitly
derive the results in the case of two noncommutative dimensions,
the generalization to arbitrary dimensionality being obvious.

It is convenient to rewrite the matrix coordinates, which satisfy
$[\hat{x}_1,\hat{x}_2]=i\theta{\rlap{1} \hskip 1.6pt \hbox{1}}$,
in terms of operators satisfying $[a,a^{\dagger}]={\rlap{1} \hskip
1.6pt \hbox{1}}$
\begin{equation}
\hat{x}_{1}=\sqrt{\frac{\theta}{2}}(a^{\dagger}+a)\qquad\qquad\hat{x}_{2}=i\sqrt{\frac{\theta}{2}}(a^{\dagger}-a).
\end{equation}
In this case, the Fourier basis matrices become
\begin{eqnarray}\label{eq:osc}
& &
\exp\left(i\sqrt{\frac{\theta}{2}}(k_1+ik_2)a^\dagger+i\sqrt{\frac{\theta}{2}}(k_1-ik_2)a\right)\\
&=&
\exp\left(-\frac{\theta}{4}(k_1^2+k_2^2)\right)\exp\left(i\sqrt{\frac{\theta}{2}}(k_1+ik_2)a^\dagger\right)\exp\left(i\sqrt{\frac{\theta}{2}}(k_1-ik_2)a\right).\nonumber
\end{eqnarray}
We can now compute the matrix element $\langle
z_1|e^{ik\cdot\hat{x}}|z_2\rangle$, where
\begin{equation}\label{eq:coherent}
|z\rangle=\exp\left(-\frac{|z|}{2}^2\right)\sum_n\frac{(za^\dagger)^n}{n!}|0\rangle
\end{equation}
is the normalized eigenstate of $a$ with eigenvalue $z$. Using
Eqs.~(\ref{eq:osc}) and (\ref{eq:coherent}) we get
\begin{eqnarray}\label{eq:ele}
\langle z_1|e^{ik\cdot\hat{x}}|z_2\rangle &=& e^{-\frac{\theta}{4}
k^2+i\sqrt{\frac{\theta}{2}}(k_1+ik_2)\bar{z}_1+i\sqrt{\frac{\theta}{2}}(k_1-ik_2)z_2}\langle
z_1|z_2\rangle\\
\langle z_1|z_2\rangle &=&
e^{-\frac{1}{2}|z_1|^2-\frac{1}{2}|z_2|^2+\bar{z}_1z_2}.
\end{eqnarray}

By performing the integrals in radial coordinates defined by
$z=re^{i\phi}$, it is easy to show that the trace can be expressed
as
\begin{equation}\label{eq:trace}
{\rm Tr}\left(e^{ik\cdot\hat{x}}\right)=\int\frac{{\rm d}z{\rm
d}\bar{z}}{2\pi i}\langle z|e^{ik\cdot\hat{x}}|z\rangle .
\end{equation}
Using Eq.~(\ref{eq:ele}) and rewriting
$z=\frac{1}{\sqrt{2\theta}}(x_1+ix_2)$, we find
Eq.~(\ref{eq:trace}) is equal to
\begin{equation}
\int\frac{{\rm d}^2x}{2\pi\theta}e^{ik\cdot x-\frac{\theta}{4}
k^2}=\frac{(2\pi)^2\delta^2(k)}{2\pi\theta},
\end{equation}
which can be easily generalized to the $2p$-dimensional case by
taking $p$ tensor products of independent two-planes. In the end,
we arrive at Eq.~(\ref{eq:ortho}). Again using Eq.~(\ref{eq:ele}),
it is straight forward to show that
\begin{equation}\label{eq:tensor}
\int\frac{{\rm d}^2k}{(2\pi)^2}\langle
z_1|e^{ik\cdot\hat{x}}|z_2\rangle \langle
z_3|e^{-ik\cdot\hat{x}}|z_4\rangle=\frac{\langle
z_1|z_4\rangle\langle z_3|z_2\rangle}{2\pi\theta}
\end{equation}
by performing the gaussian $k$ integrals. The fact that
Eq.~(\ref{eq:tensor}) holds for arbitrary $z_1$, $z_2$, $z_3$, and
$z_4$ implies Eq.~(\ref{eq:comp}) after generalizing to $2p$
dimensions.

\chapter{Propagator in the Fourier Matrix Basis}\label{ap:prop}
\index{Appendix!My Appendix \#2@\emph{My Appendix \#2}}%

In order to clarify the field theory interpretation of the matrix
propagator, we seek a representation of the form
\begin{equation}
\frac{1}{\omega^2-M^2}=\int\frac{{\rm
d}^{2p}k}{(2\pi)^{2p}}e^{-ik\cdot(B\otimes 1-1\otimes
B)}\tilde{c}(k).
\end{equation}
The Fourier coefficients, $\tilde{c}(k)$, can be constrained by
acting with $\omega^2-M^2$
\begin{eqnarray}
  1 &=& \int\frac{{\rm d}^{2p}k}{(2\pi)^{2p}}\left(\omega^2-M^2\right)e^{-ik\cdot(B\otimes 1-1\otimes B)}\tilde{c}(k)\\
 &=& \int\frac{{\rm d}^{2p}k}{(2\pi)^{2p}}\left(\omega^2+\partial_k^2\right)e^{-ik\cdot(B\otimes 1-1\otimes B)}\tilde{c}(k)+\ldots ,\nonumber
\end{eqnarray}
where the $\ldots$ represent commutator terms that are necessary
to resolve the ordering of the noncommuting matrices, $B^i\otimes
{\rlap{1} \hskip 1.6pt \hbox{1}}-{\rlap{1} \hskip 1.6pt
\hbox{1}}\otimes B^i$. It is easy to see that the commutator
corrections are negligible if
\begin{equation}
B^i\gg \left[k\cdot B,B^i\right], \left[k\cdot B,\left[k\cdot
B,B^i\right]\right],\ldots
\end{equation}
Using Eqs.~(\ref{eq:defx}) and (\ref{eq:cov}) along with the
expression for $B^i$, we see that $[k\cdot
B,\quad]=k\cdot\theta\cdot D$ where $D_i$ is the gauge covariant
derivative. Therefore the commutators are small if $\theta\cdot
k\ll L$, $L$ being the length scale set by the curvature of the
background.

Assuming that the commutators are negligible, we may keep only the
leading term in above equation. Then, upon an integration by
parts, the condition on the fourier coefficients becomes
\begin{equation}
\left(\omega^2+\partial_k^2\right)\tilde{c}(k)=(2\pi)^{2p}\delta^{2p}(k).
\end{equation}
From this equation, we arrive at the integral expression
\begin{equation}
\tilde{c}(k)=\int {\rm d}^{2p}x\frac{e^{ik\cdot
(x-x^\prime)}}{\omega^2-(x-x^\prime)^2}.
\end{equation}
Note that the consistency condition $\theta\cdot k\ll L$ can be
implemented by imposing $|x-x^\prime|\gg \theta /L$. As we will
see in Section~\ref{subsec:prop}, there is a physical reason to
impose the cutoff on the $x$ integral rather than the $k$
integral. Therefore, we will apply a cutoff of the form
$|x-x^\prime|>\theta\Lambda\gg \theta /L$. Putting everything
together, up to commutator terms that are suppressed by factors of
$(L\Lambda)^{-1}\ll 1$, we obtain the desired representation given
by Eq.~(\ref{eq:rep}).

\chapter{Equivalence Between the Fourier Basis Matrices and Wilson
Lines}\label{ap:line}
\index{Appendix!My Appendix \#3@\emph{My Appendix \#3}}%

In this Appendix, we shall verify that the Fourier basis matrices
$e^{ik\cdot\hat{x}}$ correspond to Wilson lines, which was first
obtained in \cite{ni:obse}. We begin by rewriting
\begin{equation}
e^{ik\cdot B}=\left(e^{\frac{i}{N}k\cdot B}\right)^N.
\end{equation}
Substituting in the expression for the background field
$B^i=\hat{x}^{i}+\theta^{ij}A_j(\hat{x})$, we obtain
\begin{equation}
\left(e^{\frac{i}{N}k\cdot\hat{x}+\frac{i}{N}k\cdot\theta\cdot
A(\hat{x})}\right)^N=\left(e^{\frac{i}{N}k\cdot\hat{x}}e^{\frac{i}{N}k\cdot\theta\cdot
A(\hat{x})+\ldots}\right)^N,
\end{equation}
where the $\ldots$ represent higher derivatives of the gauge field
given by powers of $\frac{1}{N}k\cdot\theta\cdot\partial$ acting
on $k\cdot\theta\cdot A(\hat{x})$. However, for arbitrary
$A(\hat{x})$ satisfying the physical boundary condition that the
field configuration vanish at infinity, we may choose $N$ large
enough to suppress the higher derivative terms. In particular, we
have
\begin{equation}\label{eq:noderiv}
e^{ik\cdot
B}=\lim_{N\rightarrow\infty}\left(e^{\frac{i}{N}k\cdot\hat{x}}e^{\frac{i}{N}k\cdot\theta\cdot
A(\hat{x})}\right)^N.
\end{equation}
We can now expand the product as follows
\begin{equation}\label{eq:expand}
\prod_{n=1}^N\left(e^{i\frac{n}{N}k\cdot\hat{x}}e^{\frac{i}{N}k\cdot\theta\cdot
A(\hat{x})}e^{-i\frac{n}{N}k\cdot\hat{x}}\right)e^{ik\cdot\hat{x}}=\prod_{n=1}^N\left(e^{\frac{i}{N}k\cdot\theta\cdot
A\left(\hat{x}+\frac{n}{N}\theta\cdot
k\right)}\right)e^{ik\cdot\hat{x}}.
\end{equation}
Again, we use the fact that physical field configurations must
vanish at infinity, so that in the $N\rightarrow\infty$ limit we
may keep only up to order $1/N$ in each exponential factor
\begin{equation}\label{eq:noexp}
\lim_{N\rightarrow\infty}\prod_{n=1}^N\left(e^{\frac{i}{N}k\cdot\theta\cdot
A\left(\hat{x}+\frac{n}{N}\theta\cdot
k\right)}\right)=\lim_{N\rightarrow\infty}\prod_{n=1}^N\Big(1+\frac{i}{N}k\cdot\theta\cdot
A\left(\hat{x}+\frac{n}{N}\theta\cdot k\right)\Big).
\end{equation}
It is now convenient to introduce the discrete path ordered
exponential
\begin{eqnarray}\label{eq:path}
& & P\exp_N\left(\sum_{n=1}^N\frac{i}{N}k\cdot\theta\cdot
A\left(\hat{x}+\frac{n}{N}\theta\cdot k\right)\right)\equiv\\
& & 1+\sum_{J=1}^N\sum_{n_1< \cdots
<n_J}\underbrace{\frac{i}{N}k\cdot\theta\cdot
A\left(\hat{x}+\frac{n_1}{N}\theta\cdot
k\right)\cdots\frac{i}{N}k\cdot\theta\cdot
A\left(\hat{x}+\frac{n_J}{N}\theta\cdot
k\right)}_{J\,\textrm{factors}},\nonumber
\end{eqnarray}
which is defined as the finite sum of $N+1$ terms, as in
Eq.~(\ref{eq:path}) above, and approaches the standard path
ordered exponential in the $N\rightarrow\infty$ limit. By using
Eqs.~(\ref{eq:noderiv}), (\ref{eq:expand}), (\ref{eq:noexp}), and
(\ref{eq:path}) and taking the continuum limit, we have
\begin{equation}
e^{ik\cdot B}=Pe^{i\int{\rm d}\sigma k\cdot\theta\cdot
A(\hat{x}+\sigma\theta\cdot k)}e^{ik\cdot\hat{x}}.
\end{equation}
Finally, using the correspondence defined by Eqs.~(\ref{eq:co})
and (\ref{eq:prod}) we arrive at Eq.~(\ref{eq:line}).

\chapter{Next to Leading Order One-Loop
Diagram}\label{ap:one-loop}
\index{Appendix!My Appendix \#3@\emph{My Appendix \#3}}%

As discussed in Section~\ref{subsec:one}, the structure of the
next to leading order one-loop contribution can be obtained by
computing
\begin{center}
\includegraphics[scale=0.25]{2_insertion.eps}.
\end{center}
A straight forward second order perturbative treatment of the
interaction term in $L_2$ that is proportional to $[B^i,B^j]$
yields a quantity of the form
\begin{eqnarray}
& & \int{\rm d}t_1{\rm d}^{2p}x_1{\rm d}t_2{\rm
d}^{2p}x_2\int\frac{{\rm d}\omega_1 {\rm
d}^{2p}k_1}{2\pi(2\pi)^{2p}}\frac{{\rm d}\omega_2 {\rm
d}^{2p}k_2}{2\pi(2\pi)^{2p}}e^{-i\omega_1(t_1-t_2)+ik_1\cdot
x_1}e^{-i\omega_2(t_2-t_1)+ik_2\cdot x_2}\nonumber\\
& & \qquad\times \widetilde{G}(\omega_1,\theta^{-1}x_1)\widetilde{G}({\omega}_2,\theta^{-1}x_2)\bigg[c_1{\rm Tr}\Big(\left[B^i,B^j\right](t_1)e^{ik_1\cdot B(t_1)}\left[B^i,B^j\right](t_2)\nonumber\\
& & \qquad\qquad\times e^{-ik_2\cdot B(t_2)}\Big){\rm
Tr}\Big(e^{-ik_1\cdot B(t_1)}e^{ik_2\cdot B(t_2)}\Big)+c_2{\rm
Tr}\Big(\left[B^i,B^j\right](t_1)e^{ik_1\cdot
B(t_1)}\nonumber\\
& & \qquad\qquad\qquad\times e^{-ik_2\cdot B(t_2)}\Big){\rm
Tr}\Big(\left[B^i,B^j\right](t_2)e^{-ik_1\cdot B(t_1)}e^{ik_2\cdot
B(t_2)}\Big)\bigg].
\end{eqnarray}
Since we only integrate out virtual states with high energy and
momentum, time derivatives of the background as well as higher
commutator terms are further suppressed. Therefore, to lowest
order, we obtain
\begin{eqnarray}
& & \int{\rm d}t{\rm d}^{2p}x_1 {\rm d}^{2p}x_2\int\frac{{\rm
d}\omega}{2\pi}\frac{ {\rm d}^{2p}k_1}{(2\pi)^{2p}}\frac{{\rm
d}^{2p}k_2}{(2\pi)^{2p}}e^{ik_1\cdot x_1}e^{ik_2\cdot x_2}
\widetilde{G}(\omega,\theta^{-1}x_1)\widetilde{G}(\omega,\theta^{-1}x_2)\nonumber\\
& & \qquad\times\bigg[c_1{\rm
Tr}\Big(\left[B^i,B^j\right](t)\left[B^i,B^j\right](t)e^{i(k_1-k_2)\cdot
B(t)}\Big){\rm Tr}\Big(e^{-i(k_1-k_2)\cdot
B(t)}\Big)\nonumber\\
& & \qquad\qquad +c_2{\rm
Tr}\Big(\left[B^i,B^j\right](t)e^{i(k_1-k_2)\cdot B(t)}\Big){\rm
Tr}\Big(\left[B^i,B^j\right](t)e^{-i(k_1-k_2)\cdot
B(t)}\Big)\bigg].\nonumber\\
\end{eqnarray}
Note that the global $SO(2p,1)$ symmetry of Eq.~(\ref{eq:mat})
requires that the time derivatives appear in the combination given
by $\dot{B}^{i2}+[B^i,B^j]^2$. Upon Fourier transforming to
position space, we are finally left with Eq.~(\ref{eq:ntlo}).

\chapter{Leading Order Two-Loop Diagrams}\label{ap:twoloop}
\index{Appendix!My Appendix \#3@\emph{My Appendix \#3}}%

The leading two-loop diagrams come from the first order
perturbative treatment of the quartic interaction terms in $L_4$
and the second order perturbative treatment of the cubic
interaction terms in $L_3$. A straight forward evaluation of the
relevant matrix elements gives quantities proportional to
{\setlength\arraycolsep{2pt}
\begin{eqnarray}\label{eq:twoloop1}
& & \int{\rm d}t{\rm d}^{2p}x_1{\rm d}^{2p}x_2\int\frac{{\rm
d}\omega_1{\rm d}^{2p}k_1}{2\pi(2\pi)^{2p}}\frac{{\rm d}\omega_2
{\rm d}^{2p}k_2}{2\pi(2\pi)^{2p}}e^{ik_1\cdot x_1}e^{ik_2\cdot x_2}\widetilde{G}(\omega_1,\theta^{-1}x_1)\widetilde{G}(\omega_2,\theta^{-1}x_2) \nonumber\\
& & \qquad\qquad \times \bigg[{\rm Tr}\Big(e^{ik_1\cdot
B(t)}\Big){\rm Tr}\Big(e^{-ik_2\cdot B(t)}\Big){\rm
Tr}\Big(e^{-ik_1\cdot B(t)}e^{ik_2\cdot B(t)}\Big) \nonumber\\
& & \qquad\qquad\qquad\qquad -{\rm Tr}\Big(e^{ik_1\cdot
B(t)}e^{ik_2\cdot B(t)}e^{-ik_1\cdot B(t)}e^{-ik_2\cdot
B(t)}\Big)\bigg],
\end{eqnarray}}\\
in the case of the quartic interactions and
\begin{eqnarray}\label{eq:twoloop2}
& & \int{\rm d}t_1{\rm d}t_2{\rm d}^{2p}x_1{\rm d}^{2p}x_2{\rm
d}^{2p}x_3 \int \frac{{\rm d}\omega_1 {\rm
d}^{2p}k_1}{2\pi(2\pi)^{2p}}\frac{{\rm d}\omega_2 {\rm
d}^{2p}k_2}{2\pi(2\pi)^{2p}} \frac{{\rm d}\omega_3 {\rm
d}^{2p}k_3}{2\pi(2\pi)^{2p}}e^{-i\omega_1(t_1-t_2)+ik_1\cdot
x_1}\nonumber\\
& & \qquad\times  e^{-i\omega_2(t_1-t_2)+ik_2\cdot
x_2}e^{-i\omega_3(t_1-t_2)+ik_3\cdot
x_3}\widetilde{G}(\omega_1,\theta^{-1}x_1)
\widetilde{G}(\omega_2,\theta^{-1}x_2)\nonumber\\
& & \qquad\times \widetilde{G}(\omega_3,\theta^{-1}x_3)
\left(\omega_1\omega_2+ \frac{\partial}{\partial
k_2}\cdot\frac{\partial}{\partial
k_3}\right)\bigg[\rm{Tr}\Big(e^{ik_1\cdot B(t_1)}e^{ik_2\cdot
B(t_1)}\Big)\rm{Tr}\Big(e^{-ik_1\cdot
B(t_1)}\nonumber\\
& & \qquad\times e^{ik_3\cdot
B(t_2)}\Big)\rm{Tr}\Big(e^{-ik_2\cdot B(t_1)}e^{-ik_3\cdot
B(t_2)}\Big)-\rm{Tr}\Big(e^{ik_1\cdot B(t_1)} e^{-ik_3\cdot
B(t_2)}e^{-ik_2\cdot B(t_1)}\nonumber\\
& & \qquad \times e^{-ik_1\cdot B(t_1)}e^{ik_3\cdot
B(t_2)}e^{ik_2\cdot B(t_1)}\Big)\bigg],
\end{eqnarray}
in the case of the cubic interactions. The structure of the triple
trace terms is by now familiar, and they correspond to the planar
matrix diagrams
\begin{center}
\begin{minipage}[c]{1.75cm}
\includegraphics[scale=0.3]{cubic.eps}
\end{minipage}%
\begin{minipage}[c]{0.5cm}
$+$
\end{minipage}%
\begin{minipage}[c]{2.7cm}
\includegraphics[scale=0.4]{quartic.eps}.
\end{minipage}
\end{center}
Using by now familiar techniques, we neglect the higher order
commutators and time derivatives in order to obtain
Eq.~(\ref{eq:threebody}) from Eq.~(\ref{eq:twoloop1}) and
Eq.~(\ref{eq:cubicthreebody}) from  Eq.~(\ref{eq:twoloop2}), after
a Fourier transformation to position space.

The single trace terms, on the other hand, are qualitatively new,
and they correspond to the nonplanar matrix diagrams
\begin{center}
\begin{minipage}[c]{1.75cm}
\includegraphics[scale=0.4]{cubic_non_planar.eps}
\end{minipage}%
\begin{minipage}[c]{0.5cm}
$+$
\end{minipage}%
\begin{minipage}[c]{2.7cm}
\includegraphics[scale=0.4]{non_planar_matrix.eps}.
\end{minipage}
\end{center}
Let us take some time to understand the structure of these
contributions. In the case of the single trace term from
Eq.~(\ref{eq:twoloop1}), using Eq.~(\ref{eq:line}), we can
simplify the trace as follows
\begin{eqnarray}\label{eq:nptrace}
& & {\rm Tr}\left(e^{ik_1\cdot B}e^{ik_2\cdot B}e^{-ik_1\cdot
B}e^{-ik_2\cdot B}\right)\nonumber\\
&=& e^{ik_1\cdot\theta\cdot k_2}\int{\rm d}x_3{\rm
tr}_N\left(P_\star e^{i\oint_C {\rm d}\gamma\cdot
A(x_3+\gamma)}\right),
\end{eqnarray}
where the contour $C=C(k_1,k_2)$ is the parallelogram defined by
the vectors $\theta\cdot k_1$ and $\theta\cdot k_2$. Using similar
techniques, it is easy to see that the single trace term in
Eq.~(\ref{eq:twoloop2}) reduces to a Wilson loop defined by a
hexagonal contour. Thus, in the case of nonplanar matrix diagrams,
the gauge invariant contribution of background gauge field
insertions can form Wilson loops, in addition to Wilson lines.

We can shed further light on the meaning of nonplanar matrix
diagrams by considering the vacuum diagram associated with the
single trace term of Eq.~(\ref{eq:twoloop1}), which is obtained by
setting $A_i=0$.  This leaves a quantity proportional to
\begin{eqnarray}\label{eq:vac}
& & NV\int{\rm d}t{\rm d}^{2p}x_1{\rm d}^{2p}x_2\int\frac{{\rm
d}\omega_1{\rm d}^{2p}k_1}{2\pi(2\pi)^{2p}}\frac{{\rm d}\omega_2
{\rm d}^{2p}k_2}{2\pi(2\pi)^{2p}}e^{ik_1\cdot x_1}e^{ik_2\cdot x_2} \nonumber\\
& & \qquad\qquad \times e^{ik_1\cdot\theta\cdot
k_2}\widetilde{G}({\omega}_1,{\theta}^{-1}x_1)\widetilde{G}({\omega}_2,{\theta}^{-1}x_2).
\end{eqnarray}
Upon performing the $k$ integrals, we are left with
\begin{eqnarray}\label{eq:exp}
 NV\int{\rm d}t\frac{{\rm d}\omega_1{\rm d}^{2p}p_1}{2\pi(2\pi)^{2p}} \frac{{\rm d}\omega_2{\rm d}^{2p}p_2} {2\pi(2\pi)^{2p}} e^{ip_1\cdot\theta\cdot p_2} \widetilde{G}(\omega_1,p_1)\widetilde{G}(\omega_2,p_2),
\end{eqnarray}
where we have changed variables of integration to $p={\theta}^{-1}
x$. Eq.~(\ref{eq:exp}) is easily recognized as the contribution to
the effective action from the two-loop nonplanar field theory
vacuum diagram
\begin{center}
\includegraphics[scale=0.5]{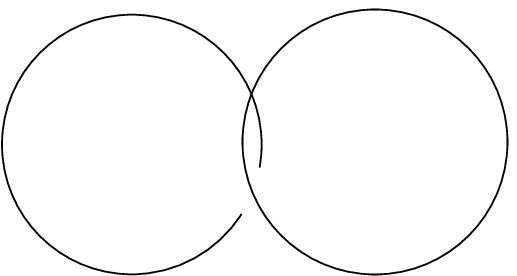}.
\end{center}
Thus, nonplanar matrix diagrams correspond to field theory
diagrams with the loops linked in a nonplanar fashion.

However, when $\theta\Lambda^2\gg 1$, nonplanar loop diagrams do
not contribute to the Wilsonian integration.  To see this,
consider the integration over $\omega,p$ in Eq.~(\ref{eq:exp})
\begin{eqnarray}
 & & \int_{\Lambda}\frac{{\rm d}\omega_1{\rm d}^{2p}p_1}{2\pi(2\pi)^{2p}}\frac{{\rm d}\omega_2{\rm d}^{2p}p_2}{2\pi(2\pi)^{2p}} e^{ip_1\cdot\theta\cdot p_2}\widetilde{G}(\omega_1,p_1)\widetilde{G}(\omega_2,p_2)\nonumber\\
 &\sim & \int_{\Lambda} \frac{{\rm d}^{2p}p_1}{|\theta p_1|}\frac{{\rm d}^{2p}p_2}{|\theta p_2|} e^{ip_1\cdot\theta\cdot p_2}\sim
 {\Lambda}^{4p-2}e^{-\theta{\Lambda}^{2}},
\end{eqnarray}
where the final integration can be performed with Schwinger
parameters in the stationary phase approximation. Thus, in the UV
domain of Wilsonian integration, the contribution from nonplanar
matrix diagrams is exponentially suppressed due to Moyal phase
factors and, therefore, negligible compared to planar matrix
diagrams. In fact, we will restrict our analysis to the
$\theta\Lambda^2\gg 1$ regime in order to avoid nonplanarity. The
reason is that, according to the intuition that we have so far
developed, each trace is associated with a point in space, and
therefore, the reduction in the number of traces that occurs in
nonplanar matrix diagrams is most naturally interpreted as some
kind of short-distance effect. Since the intuitive picture of UV
dipoles propagating in loops that we discussed in
Section~\ref{subsec:one} is not compatible with the notion
short-distance quantum corrections, we regard the nonplanar
contributions that emerge when $\theta\Lambda^2\lesssim 1$ as
indicating the presence of different degrees of freedom in the IR
-- possibly Wilson loop in nature.

Nonetheless, nonplanar matrix diagrams can, in principle, be
treated in the context of the background derivative expansion. As
usual, one thinks of the $k$ variables as small, and therefore,
expands the Wilson loop. This provides the insertion of higher
dimensional operators into the trace as well as higher powers of
$k$, which are equivalent to more powers of momentum in the
denominator by means of integration by parts. However, one cannot
expand the $\exp(k_1\cdot\theta\cdot k_2)$ factor, since it leads
to the Moyal phase factor that ultimately suppresses the
contribution.

\chapter{Next to Leading Order Two-Loop Diagrams}\label{ap:ntlotl}
\index{Appendix!My Appendix \#3@\emph{My Appendix \#3}}%

As in the next to leading order one-loop calculation, to get the
precise result we must retain higher order commutators and time
derivatives that were dropped in the derivation of the matrix
propagator Eq.~(\ref{eq:prop}) as well as the field strength terms
in $L_2$ that were also excluded from the the propagator. However,
after already developing some intuition  for the structure of
terms that can appear, a lengthly calculation is not necessary. We
simply need to keep track of all the distinct possibilities in
which the field strength insertions can appear in the diagrams.
For example, in the case of the quartic diagram, both insertions
can go into one loop or each loop can get a single insertion
\begin{center}
\begin{minipage}[c]{2.7cm}
\includegraphics[scale=0.4]{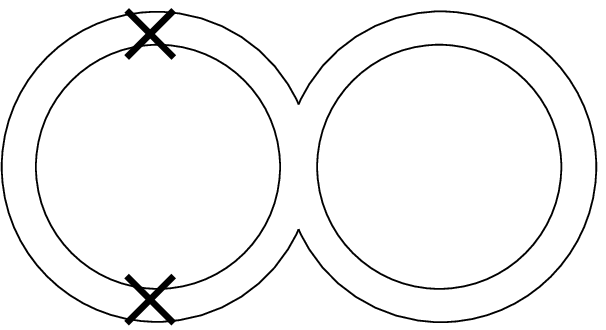}
\end{minipage}%
\begin{minipage}[c]{0.5cm}
or
\end{minipage}%
\begin{minipage}[c]{2.7cm}
\includegraphics[scale=0.4]{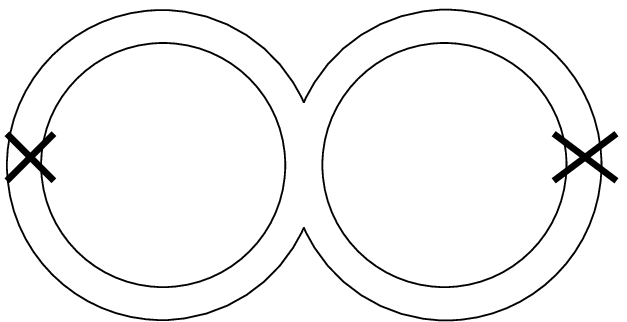}.
\end{minipage}
\end{center}
Of course, each insertion requires an extra field theory
propagator to appear in the corresponding loop, and for each loop,
there are two boundaries in which the insertion can go. Therefore,
the first diagram corresponding to both insertions going into the
same loop contributes two terms
\begin{eqnarray}\label{eq:leading}
& & \int{\rm d}t{\rm d}^{2p}x_1{\rm d}^{2p}x_2{\rm d}^{2p}x_3\int
{\rm d}\omega_1{\rm d}\omega_2\widetilde{G}(\omega_1,\theta^{-1}x_{13})^{3}\widetilde{G}(\omega_2,\theta^{-1}x_{23})\nonumber\\
& &
\qquad\times\Big[b_1\rho_{FF}(x_1,t)\rho(x_2,t)\rho(x_3,t)+b_2\rho_F(x_1,t)\rho(x_2,t)\rho_F(x_3,t)\Big],
\end{eqnarray}
while the second diagram corresponding to one insertion going into
each loop contributes three terms
\begin{eqnarray}\label{eq:beta}
& & \int{\rm d}t{\rm d}^{2p}x_1{\rm d}^{2p}x_2{\rm d}^{2p}x_3\int
{\rm d}\omega_1{\rm d}\omega_2\widetilde{G}(\omega_1,\theta^{-1}x_{13})^{2}\widetilde{G}(\omega_2,\theta^{-1}x_{23})^{2}\nonumber\\
& &
\qquad\times\Big[b_3\rho_F(x_1,t)\rho(x_2,t)_F\rho(x_3,t)+b_4\rho(x_1,t)\rho(x_2,t)\rho_{FF}(x_3,t)\nonumber\\
& & \qquad\qquad\qquad\qquad\qquad\qquad
+b_5\rho_F(x_1,t)\rho(x_2,t)\rho_F(x_3,t)\Big].
\end{eqnarray}

There are many more possibilities in the case of the cubic graph.
For simplicity, we will enumerate them in four propagator form,
which is obtained after the cancellation of one propagator by the
momenta in the numerator of the integrand. There are three
combinations of field theory propagators that emerge: the first
can be included in Eq.~(\ref{eq:leading}), the second can be
included in Eq.~(\ref{eq:beta}), and the third is given by
\begin{eqnarray}\label{eq:dangerous}
& & \int{\rm d}t{\rm d}^{2p}x_1{\rm d}^{2p}x_2{\rm d}^{2p}x_3\int{\rm d}\omega_1{\rm d}\omega_2\widetilde{G}(\omega_1,\theta^{-1}x_{12})^{2}\widetilde{G}(\omega_2,\theta^{-1}x_{23})\nonumber\\
& &
\qquad\times\widetilde{G}(\omega_1+\omega_2,\theta^{-1}x_{13})\Big[b_6\rho_{FF}(x_1,t)\rho(x_2,t)\rho(x_3,t)+b_7\rho_F(x_1,t)\rho_F(x_2,t)\nonumber\\
& & \qquad\qquad\qquad\qquad\qquad\qquad
\times\rho(x_3,t)+b_8\rho_F(x_1,t)\rho(x_2,t)\rho_F(x_3,t)\Big].
\end{eqnarray}
Thus, we can parameterize the inequivalent next to leading order
two loop contributions by proportionality constants $b_1$ through
$b_8$. Note that by focusing on the constant part of $\rho$, as we
did in the one-loop next to leading order calculation, we obtain
two-loop corrections to Eq.~(\ref{eq:ntlo}). In particular, the
two-loop contribution to the beta function is contained in
Eq.~(\ref{eq:dangerous}).

\begin{vita}
Eric Alexander Nicholson was born in Hays, Kansas on 17 December
1975, the son of Jack A. Nicholson and Nancy J. Mahr.  He received
the Bachelor of Science degree in Physics from the Massachusetts
Institute of Technology in 1998. He then moved on to the
University of Texas at Austin for his graduate studies until 2003.
Upon receiving his Ph.D., he plans to join the Department of
Defense as an Operations Research Analyst.

\end{vita}


\begin{thebibliography}{99}


\bibitem{ib:twotime} I.~Bars and C.~Deliduman, Phys. Rev. D \textbf{58}, 066004
(1998); I.~Bars, C.~Deliduman and D.~Minic, Phys. Lett. B
\textbf{457}, 275 (1999).
\bibitem{sw:stri} N.~Seiberg and E.~Witten, J. High Energy Phys. \textbf{09},
032 (1999).
\bibitem{sb:magn} M.~M.~Sheikh-Jabbari, Phys. Lett. B \textbf{455}, 129 (1999);
D.~Bigatti and L.~Susskind, Phys. Rev. D \textbf{62}, 066004
(2000); Z.~Yin, Phys. Lett. B \textbf{466}, 234 (1999); H.~Liu and
J.~Michelson, Phys. Rev. D \textbf{62}, 066003 (2000).
\bibitem{mr:uvdiv} C.P.~Martin and D.~Sanchez-Ruiz, Phys. Rev.
Lett. \textbf{83}, 476 (1999); A.~Armoni, Nucl. Phys.
\textbf{B593}, 229 (2001).
\bibitem{cm:dual} C.P.~Martin and F.R.~Ruiz, Nucl. Phys.
\textbf{B597}, 197 (2001).
\bibitem{cr:renorm}  T.~Krajewski and R.~Wulkenhaar,
Int. J. Mod. Phys. A \textbf{15}, 1011 (2000); I.~Chepelev and
R.~Roiban, J. High Energy Phys. \textbf{05}, 037 (2000);
H.~Girotti, M.~Gomes, V.~Rivelles and A.~da~Silva, Nucl. Phys.
\textbf{B587}, 299 (2000); A.~Bichl, \textit{et al}, J. High
Energy Phys. \textbf{10}, 046 (2000); I.~Chepelev and R.~Roiban,
\emph{ibid}. \textbf{03}, 001 (2001); A.~Bichl, \textit{et al},
\emph{ibid}. \textbf{06}, 013 (2001); S.~Sarkar, \emph{ibid}.
\textbf{06}, 003 (2002).
\bibitem{ms:nonp} S.~Minwalla, M.~Van Raamsdonk and N.~Seiberg, J. High Energy Phys. \textbf{02}, 020
(2000).
\bibitem{mh:irdiv} M.~Hayakawa, Phys. Lett. B \textbf{478}, 394 (2000); F.R.~Ruiz, \emph{ibid}. \textbf{502}, 274 (2001).
\bibitem{gp:wilsonrg} L.~Griguolo and M.~Pietroni, J. High Energy
Phys. \textbf{05}, 032 (2001); L.~Griguolo and M.~Pietroni, Phys.
Rev. Lett. \textbf{88}, 071601 (2002); G.H.~Chen and Y.S.~Yu,
Nucl. Phys. \textbf{B622}, 189 (2002).
\bibitem{vv:wilson}  V.~V.~Khoze and G.~Travaglini, J. High Energy Phys. \textbf{01}, 026
(2001); T.~J.~Hollowood, V.~V.~Khoze and G.~Travaglini,
\emph{ibid}. \textbf{05}, 051 (2001); C.~S.~Chu,  V.~V.~Khoze and
G.~Travaglini, Phys. Lett. B \textbf{513}, 200 (2001); C.~S.~Chu,
V.~V.~Khoze and G.~Travaglini, \emph{ibid}. \textbf{543}, 318
(2002).
\bibitem{ls:gaug} A.~Matusis, L.~Susskind and N.~Toumbas, J. High Energy Phys.
\textbf{12}, 002 (2000); D.~Zanon, Phys. Lett. B \textbf{502}, 265
(2001); M.~Pernici, A.~Santambrogio and D.~Zanon, \emph{ibid}.
\textbf{504}, 131 (2001); A.~Santambrogio and D.~Zanon, J. High
Energy Phys. \textbf{01}, 024 (2001).
\bibitem{ki:inte} Y.~Kiem, S.~Lee, S.~J.~Rey and H.~T.~Sato, Phys. Rev. D \textbf{65}, 046003 (2002);
Y.~Kiem, S.~J.~Rey, H.~T.~Sato and J.~T.~Yee, \emph{ibid}.
\textbf{65}, 026002 (2002); Y.~Kiem, S.~Lee, S.~J.~Rey and
H.~T.~Sato, Eur. Phys. J. \textbf{C}22, 757 (2002); Y.~Kiem,
S.~S.~Kim, S.~J.~Rey and H.~T.~Sato, Nucl. Phys. \textbf{B641},
256 (2002).
\bibitem{ar:uvir} A.~Armoni and E.~Lopez, Nucl. Phys. \textbf{B632}, 240 (2002).
\bibitem{hl:trek} H.~Liu and J.~Michelson, Nucl. Phys. \textbf{B614}, 279 (2001);
H.~Liu, \textit{ibid}. \textbf{B614}, 305 (2001).
\bibitem{mv:mean} M.~Van~Raamsdonk, J. High Energy Phys. \textbf{11}, 006 (2001).
\bibitem{jn:dipo} L.~Jiang and E.~Nicholson, Phys. Rev. D
\textbf{65}, 105020 (2002).
\bibitem{en:renorm} E.~Nicholson, Phys. Rev. D \textbf{66}, 105018
(2002).
\bibitem{kk:bilocal} S.~Iso, H.~Kawai and Y.~Kitazawa, Nucl.
Phys. \textbf{B576}, 375 (2000).
\bibitem{rg:soli} R.~Gopakumar, S.~Minwalla and A.~Strominger, J. High Energy Phys. \textbf{05}, 020 (2000).
\bibitem{ns:back} N.~Seiberg, J. High Energy Phys. \textbf{09}, 003 (2000).
\bibitem{ni:obse} N.~Ishibashi, S.~Iso, H.~Kawai and Y.~Kitazawa, Nucl. Phys. \textbf{B573}, 573 (2000);
S.~J.~Rey and R.~von~Unge, Phys. Lett. B \textbf{499}, 215 (2001);
S.~Das and S.~J.~Rey, Nucl. Phys. \textbf{B590}, 453 (2000);
D.~Gross, A.~Hashimoto and N.~Itzhaki, Adv. Theor. Math. Phys.
\textbf{4}, 893 (2000).
\bibitem{rg:ncos} R.~Gopakumar, S.~Minwalla, N.~Seiberg and
A.~Strominger, J. High Energy Phys. \textbf{08}, 008 (2000).
\bibitem{ar:closed} A.~Rajaraman and M.~Rozali, J. High Energy
Phys. \textbf{04}, 033 (2000).
\bibitem{sj:dual} S.~J.~Rey, hep-th/0207108; S.~J.~Rey and
J.~T.~Yee, hep-th/0303235.

\end{thebibliography}
\end{document}